\documentclass[fleqn,usenatbib]{mnras}

\usepackage{newtxtext,newtxmath}

\usepackage[T1]{fontenc}
\usepackage{lineno}
\DeclareRobustCommand{\VAN}[3]{#2}
\let\VANthebibliography\thebibliography
\def\thebibliography{\DeclareRobustCommand{\VAN}[3]{##3}\VANthebibliography}

\usepackage{graphicx}	
\usepackage{amsmath}	

\title[BNS Merger-nova LF and ToO detection efficiency]{Luminosity Functions and Detectability of Binary Neutron Star Merger-nova Signals with Various Merger Remnants}

\author[Chen et al.]{
Zhiwei Chen,$^{1,2}$
Youjun Lu,$^{1,2}$\thanks{E-mail: luyj@nao.cas.cn}
Hao Ma$^{3}$, and Qingbo Chu$^{1,2}$ 
\\
$^{1}$National Astronomical Observatories, Chinese Academy of Sciences, 20A Datun Road, Beijing 100101, China\\
$^{2}$School of Astronomy and Space Sciences, University of Chinese Academy of Sciences, 19A Yuquan Road, Beijing 100049, China\\
$^{3}$College of Physics and Electronic Engineering, Chengdu Normal University, Chengdu 611130, China
}

\date{Accepted XXX. Received YYY; in original form ZZZ}

\pubyear{\the\year{}}

\begin{document}
\label{firstpage}
\pagerange{\pageref{firstpage}--\pageref{lastpage}}
\maketitle

\begin{abstract}
With the rapid advancements in next-generation ground-based gravitational wave (GW) detectors, it is anticipated that $10^3$-$10^5$ binary neutron star (BNS) mergers per year will be detected, with a significant fraction accompanied by observable merger-nova signals through future sky surveys. Merger-novae are typically powered by the radioactive decay of heavy elements synthesized via the r-process. If the post-merger remnant is a {long-lived rapid-rotating neutron star}, the merger-nova can be significantly enhanced due to strong magnetized winds. {In this paper, we generate mock BNS merger samples using binary population synthesis model and classify their post-merger remnants--black hole (BH) and magnetar, (i.e., long-lived supramassive NS and stable NS), based on results from numerical simulations.} We then construct merger-nova radiation models to estimate their luminosity function. We find that the luminosity function may exhibit a distinctive triple-peak structure, with the relative positions and heights of these peaks depending on the equation of state (EOS) of the BNS. Furthermore, we estimate the average Target-of-Opportunity (ToO) detection efficiency $\langle f_{\rm eff} \rangle$ with the Chinese Space Station Telescope (CSST) and find that due to possible enhanced luminosity, the largest source redshift with $\langle f_{\rm eff} \rangle>0.1$ can be enlarged from $z_{\rm s}\sim 0.5$ to $z_{\rm s}\sim 1-1.5$. Besides, we also generate the detectable mass spectrum for merger-novae by $\langle f_{\rm eff}\rangle$, which may provide insights to the ToO searching strategy.
\end{abstract}




\section{Introduction}

Long before the first detection of the gravitational wave (GW) event GW170817 \citep{2017PhRvL.119p1101A} and its multiwavelength electromagnetic counterpart (EM) observation \citep[e.g., ][]{2017ApJ...848L..12A, 2017ApJ...848L..13A,2017Sci...358.1556C, 2019MNRAS.489L..91C, 2019PhRvX...9a1001A}, the binary neutron star (BNS) merger is widely believed to produce a thermal UV-optical or Infrared transient (i.e., kilonova or macronova, and thereafter merger-nova\footnote{{The thermal emission of the merger ejecta from the original work of \citet{1998ApJ...507L..59L} was first named as “macronova” due to its sub-supernova luminosity by \citet{2005astro.ph.10256K} and then named as “kilonova” due to its about $\sim 1000$ times than the nova by \citet{2010MNRAS.406.2650M}. In the work of \citet{2013ApJ...776L..40Y}, they first proposed a kilonova model for those BNS mergers with magnetar remnants in which the magnetar wind energy injection may significantly enhance the luminosity, and they used a more general word “merger-nova” to reflect a wider range of predicted luminosity than that of the traditional kilonovae. In this paper, we adopt ``merger-nova'' since one of the focus is considering about the magnetar wind energy injection.}}) \citep[e.g., ][]{1998ApJ...507L..59L, 2005astro.ph.10256K, 2010MNRAS.406.2650M,2013ApJ...775..113T,2013ApJ...775...18B}, powered by the radioactive decay of the heavy elements, such as lanthanide elements synthesized via the $r$-process in the neutron-rich material ejecta \citep{2017Natur.551...80K}. In the past few years,  {a series of studies has proposed the phenomenological model} by introducing different mass components \citep[e.g., ][]{2017ApJ...850L..37P, 2017ApJ...851L..21V, 2021MNRAS.505.1661B, 2023MNRAS.522..912Z}, for example, blue and red components to fit the light curve of the merger-nova emission, by assuming the central remnant object to be a black hole (BH) or a short-lived hypermassive neutron star (HMNS). Among all these models, the light curve of the merger-nova is mainly determined by the mass of the ejecta $m_{\rm ej}$ (typically $\sim10^{-4}-10^{-2} M_{\odot}$ \citep[e.g., ][]{2021MNRAS.505.1661B}) and their dynamical evolution, which is strongly dependent on the equation of state (EOS) of BNS. Therefore, the peak luminosity function of the merger-nova emission may vary with different EOS. 
{For example, \citet{2023MNRAS.522..912Z} (also \citealt{2023MNRAS.520.2829S}) found that the peak luminosity function is bimodal in optical bands, with the locations of the two peaks varying with the stiffness of the EOS, under the assumption that all the merger remnants are BHs.}

However, numerical simulations have shown that the post-merger product can also be a stable and rapidly rotating magnetar \citep[e.g.,][]{2013ApJ...763L..22Z, 2017PhRvL.119w1102S,2025ApJ...984L..61M}. Besides, there is strong observational evidence of short gamma-ray bursts (sGRBs) supporting the existence of rapidly rotating magnetar in the BNS merger, including the extended emission \citep[e.g., ][]{2006ApJ...643..266N, 2008MNRAS.385.1455M}, X-ray flares \citep[e.g.,][]{2005ApJ...635L.133B, 2006Natur.442.1008C} and more importantly the internal plateaus with rapid decay at the end of the plateaus \citep[e.g., ][]{2013MNRAS.430.1061R, 2018ApJ...869..155S,2023arXiv230705689S}. In this magnetar scenario, the spin-down of the remnant magnetar supplies an additional energy (normally at the scale of $10^{47}\rm erg ~s^{-1}$) through the strong magnetic wind, which may significantly affect the dynamical evolution of the ejecta and the resultant radiation of the merger-nova signals \citep[e.g., ][]{2006Sci...311.1127D, 2006MNRAS.372L..19F, 2013ApJ...771L..26G, 2013ApJ...776L..40Y}.  For example, the observation of the merger-nova signal associated with GRB130306B discovered by Hubble Space Telescope (HST) exhibits extreme luminous behavior in the F160W band \citep{2013Natur.500..547T}, i.e., $m_{\rm ab}=25.73\pm0.2$ with redshift $z_{\rm s}\sim 0.356$. This enhancement of luminosity may be interpreted as evidence of a magnetar central engine. 

{Recently, several theoretical studies have been conducted on merger-nova signals with magnetar remnants. For example, \citet{2024MNRAS.527.5166W} proposed to constrain the supramassive neutron stars (SMNSs) by their detection rate and \citet{2025ApJ...989L..41M} provided estimates on the detection horizon of single sources by various EM telescopes \cite[see also][]{Murase:2017snw}. }
{In this paper, however, we will focus on the luminosity function of merger-nova signals with various merger remnants, i.e., BH and magnetar scenarios and demonstrate their potential application as a probe to determine the EOS of BNS mergers statistically.}
In theory, the remnant type of the BNS merger is directly dependent on its mass and EOS \citep[e.g., ][]{2008PhRvD..78h4033B,2017ApJ...844L..19P, 2022ApJ...939...51M, 2022A&A...666A.174S}. Given the merger rate density of the BNS mergers, the fraction of these two remnants and the total merger-nova luminosity function in the universe may vary significantly with the EOS.  Firstly, we generate BNS merger samples from binary population synthesis and identify their post-merger remnant type by adopting results from both observation constraints and numerical simulations for different EOS. Then we apply the phenomenological radiation models for different remnant types to calculate the observed luminosity function of the merger-nova {and show that one may use it to provide information on both the merger-nova models and the EOS of the BNSs}. Finally, we estimate their Target-of-Opportunity detection efficiency distribution by the to-be-launched Chinese Space Station Telescope \citep{2019ApJ...883..203G} and the next-generation GW detector Cosmic Explorer \citep{2019BAAS...51g..35R}.

This paper is organized as follows. In Section~\ref{sec:method}, we introduce the main methodology, including the generation of the BNS merger sample (Section~\ref{sec:sample}), the identification of the post-merger remnant type (Section~\ref{sec:remnant})  the merger-nova models (Section~\ref{sec:model}), the detection efficiency estimation (Section~\ref{subsec:feff}), and the signal-to-noise ratio and localization area estimation for BNS mergers detected by GW detectors (Section~\ref{subsec:GW}). In Section~\ref{sec:results}, we present our main results. The conclusions and discussion are provided in Section~ \ref{sec:con}. Throughout the paper, we adopt the cosmological parameters as $(h_0,\Omega_{\rm m},\Omega_\Lambda)=(0.68,0.31,0.69)$ \citep{Aghanim2020}.

\section{Methodology}
\label{sec:method}

\subsection{BNS merger samples}
\label{sec:sample}

\begin{figure}
\centering
\includegraphics[width=0.9\columnwidth]{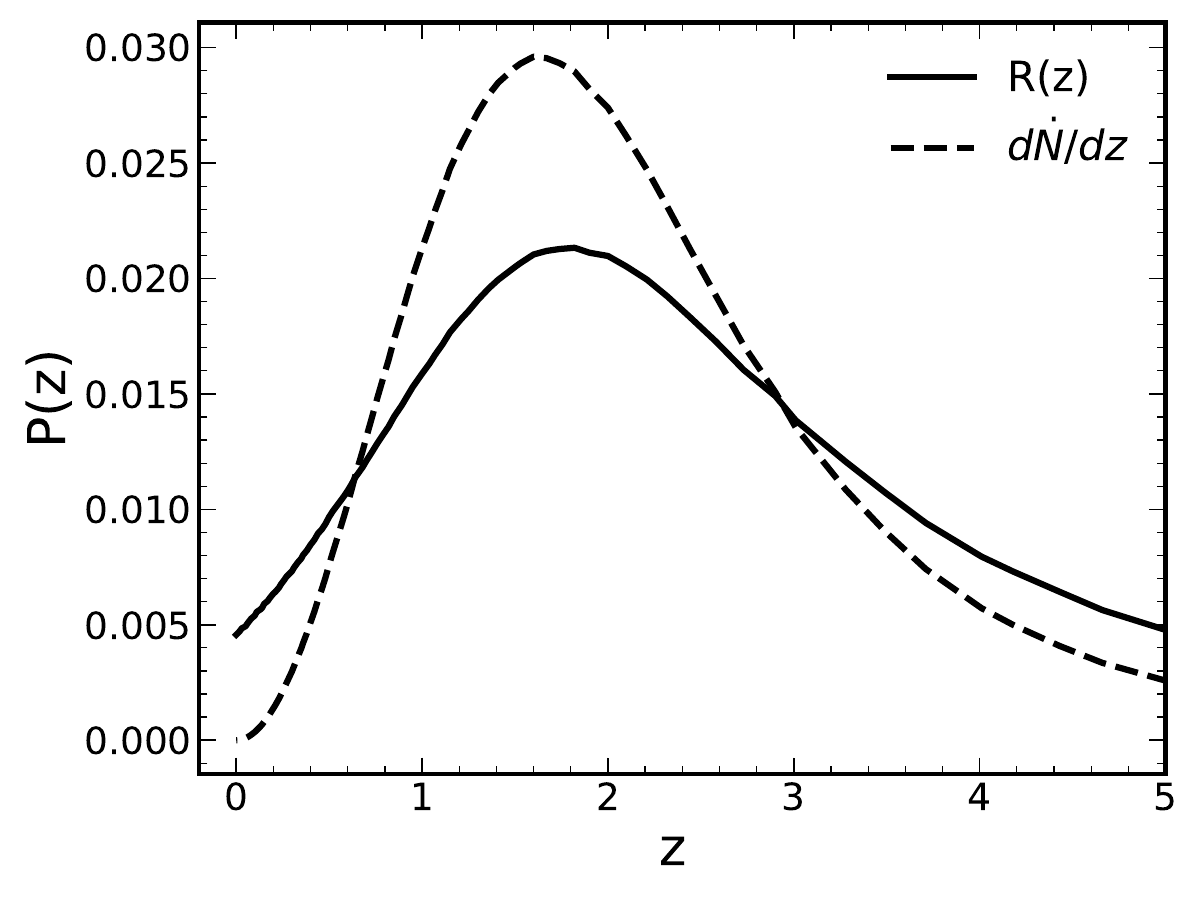}
\caption{
The normalized redshift evolution of the merger rate density, ${R}(z)$ (solid line), and the number density, $d\dot{N}/{dz}$ (dashed line), of the mock BNS merger samples (marginalized over $m_1$ and $q$) generated by the $\boldsymbol{\alpha10.\rm kb\beta0.9}$ BSE model in \citet{2022MNRAS.509.1557C} with observational extinction-corrected SFR in \citet{annurev:/content/journals/10.1146/annurev-astro-081811-125615} and the mean metallicity redshift evolution obtained in \citet{2016Natur.534..512B}. 
}
\label{fig:merger_rate}
\end{figure}

The BNS merger samples in this paper are generated by adopting the $\boldsymbol{\alpha10.\rm kb\beta0.9}$ model proposed in \citet{2022MNRAS.509.1557C}, which implements parameterized population models for binary stellar evolution (BSE) in the formation and evolution of cosmological galaxies \citep[e.g., ][]{2011MNRAS.413..101G, 2015MNRAS.446..521S, 2018MNRAS.475..648P}. In this model, the impacts of the common envelope phase, natal kick, mass ejection during the secondary SN explosion, and metallicity are taken into account. Using Bayesian methods, \citet{2022MNRAS.509.1557C} found that this model is consistent with both Galactic BNS observations and the local BNS merger rates inferred from GW observations.  The cosmic star formation rate and the metallicity evolution are taken as the same in \citet{2014ARA&A..52..415M} and \citet{2016Natur.534..512B} to obtain the number density of BNS mergers with redshift evolution. Therefore, the number distribution of BNS mergers per unit time can be written as \citep[e.g.,][]{2021MNRAS.500.1421Z, 2023ApJ...953...36C}: 
\begin{equation}
\frac{d^3\dot{N}}{d{m_1}dq dz}=\frac{{R}(z,{m_1},q)}{1+z} \frac{dV_{\rm c}(z)}{dz},  
\label{source}
\end{equation}
where $m_1$ is the primary mass, $q (=m_2/m_1)$ is the mass ratio with $m_2$ representing the secondary mass, and $R(z,m_1,q)$ is the merger rate density taken from this model with the primary mass in the range of $m_1$ to $m_1+ dm_1$, the mass ratio in the range of $q$ to $q+dq$, {the redshift in the range $z_{\rm s}$ to $z_{\rm s}+dz_{\rm s}$}. The factor $1/(1+z)$ accounts for the dilation in time. Figure~\ref{fig:merger_rate} shows the normalized redshift evolution of $R(z)$ (solid line) and $\iint \frac{d^3\dot{N}}{dm_1 dq dz} dm_1 dq$ (dashed line), respectively. As seen in this figure, the peak of the merger rate density is about $\sim 1.8$, while that of the number density is $\sim 1.6$. Both $R(z) \propto \iint R(z, m_1, q)dm_1 dq$ and $d \dot{N}/dz \propto \iint \frac{d^3\dot{N}}{dm_1 dq dz} dm_1 dq$ decay rapidly when $z>5$.

For the demonstration purpose of this paper, we choose two different EOS, namely  {DD2} \citep[e.g., ][]{2010NuPhA.837..210H, 2010PhRvC..81a5803T, 2013PhRvL.111m1101B}  {and  {SLy} \citep[e.g., ][]{2001A&A...380..151D}.} For  {DD2}, the maximum mass of non-rotating NSs, $M_{\rm TOV}$ is $2.42 M_{\odot}$, while for  {SLy} $M_{\rm TOV}=2.06M_{\odot}$. For rotating NS, the mass of $1.2 M_{\rm TOV}$ is chosen as the limit between SMNS and BH \citep[e.g., ][]{2016MNRAS.459..646B,2018ApJ...852L..25R, 2022ApJ...937...79C}, {and the radius is approximated by $R_{\rm rot}=1.34R_{\rm stat}$ \footnote{ Note that the simple empirical relation between $R_{\rm rot}$ and $R_{\rm stat}$ is accurate for the maximum-mass NSs, but not for stable NS remnants of BNS mergers with masses below $ 1~M_{\rm TOV}$. In principle, one may use the Python package  {RNS} \citep{1995ApJ...444..306S} to give a more accurate estimation of $R_{\rm rot}$ for all rotating NSs but quite time-consuming. According to calculations by \citet{2023PhRvD.108l4056K} , the radius of a rapid rotating ($f_{\rm spin}\sim 1 ~\rm kHz$) NS with $\sim 1.5 M_{\odot}$ is $\sim 13.6~\rm km$ assuming the EOS to be SLy , not substantially different from the estimation using this relation. Therefore, we adopt this relation for all rotating NSs, for simplicity.} } \citep[see][]{1996ApJ...456..300L} for the rapid-rotating case for simplicity, such as the magnetar discussed in Section~\ref{sec:remnant}. We also consider the effect of EOS on the distributions of the total mass $m_{\rm tot}$ and the mass ratio $q$ as discussed in \citet{2023MNRAS.522..912Z}. Although the merger-nova signal only slightly depends on the viewing angle, unlike afterglow signals, the viewing angle $\theta_{\rm v}$ in this paper is chosen to obey the probability distribution for a sample of BNS mergers detected by ground-based GW detectors \citep[][]{2011CQGra..28l5023S}:
\begin{equation}
P(\theta_{\rm v})=0.076(1+6\cos^2{\theta_{\rm v}}+\cos^4{\theta_{\rm v}})^{3/2}\sin{\theta_{\rm v}}. 
\end{equation}

By adopting the Gibbs Sampling method \citep[e.g., ][]{gibbs}, we randomly generate $5\times10^5$ mock BNS mergers within the redshift of $[0,5)$ with different $m_1$ and $m_2$. The radius of the NS $r_{1,2}$ is then calculated by the chosen EOS, respectively. Therefore, for each BNS merger sample, we obtain a set: $(z,m_1,m_2,r_1,r_2,\theta_{\rm v})$. By these parameters, one may directly determine the remnant types of the BNS merger samples by combining several observation constraints and GRMHD results, and therefore determine the merger-nova signals produced by these mergers. 

\subsection{Merger Remnant}
\label{sec:remnant}

The merger-nova signal of BNS mergers is strongly dependent on the merger remnant type. As discussed in several literature \citep[e.g., ][]{2008PhRvD..78h4033B,2020GReGr..52..108B,2012ApJ...746...48M, 2017ApJ...844L..19P, 2013ApJ...776L..40Y, 2019ApJ...880L..15M, 2022A&A...666A.174S}, there are three main possibilities of remnants of BNS mergers depending on their masses: 1) the merger of two NSs may directly collapse to form a BH or a short-lived HMNS, if the remnant mass is larger than $1.2 M_{\rm TOV}$.  In theory, the HMNS is supported from collapse by differential rotation rather than rigid rotation case in the SMNS and stable NS case \citep[see][for a thorough review]{2020GReGr..52..108B}. A large number of numerical simulations have shown that the lifetime of the HMNSs are short across various masses \citep[e.g., ][]{PhysRevD.95.123003,PhysRevD.98.104005,Koppel_2019}, typically around several milliseconds \citep[also see Tables 1 and 2 summarized in ][]{2020JHEAp..27...33L}. One possible explanation for such a short lifetime is the differential rotation can be damped quickly by viscous angular momentum transfer \citep{PhysRevD.97.124039,10.1093/mnras/sty2531}. Therefore, the energy injection due to the spin-down of HMNS only lasts for a short time and does not lead to the enhancement of the merger-nova signals as in the case with the BH remnant. Hereafter, we assume that the HMNS case is the same as the BH case and do not distinguish them from each other; 2) the two NSs may form a long-lived supramassive NS (hereafter SMNS) if the remnant mass is median (within the range between $1.0 M_{\rm TOV}$ and $1.2 M_{\rm TOV}$ as defined below); 3) the final remnant can be a stable NS if the remnant mass is smaller than $1.0 M_{\rm TOV}$. In this section, we outline the semi-analytical model to infer the properties of merger remnant. 

In this paper, we consider three different components of the ejecta material from a BNS merger according to different ejection mechanisms. When a BNS merges, there is a small fraction of neutron-rich matter, normally $\sim10^{-4}-10^{-2} M_{\odot}$ for the case of BH remnant case and $\sim 10^{-1} M_{\odot}$ for the NS remnant case, respectively, which can be ejected with a velocity of $\sim 0.1-0.3c$ within the dynamical timescale ($\sim \rm ms$) by the tidal forces or by shocks in the collision of the neutron star cores \citep[e.g., ][]{2017LRR....20....3M}. Numerical simulations \citep[e.g., ][]{2020PhRvD.101j3002K} have shown that the mass of the dynamical ejecta in the BH case can be approximated as
\begin{equation}
\frac{m_{\mathrm{dyn}}}{10^{-3} M_{\odot}}=\left(\frac{a}{C_1}+b \frac{m_2^n}{m_1^n}+c C_1\right) m_1+(1 \leftrightarrow 2),
\label{mass:dyn}
\end{equation}
where $C_{1,2}$ is the compactness of the two components of the BNS (1 and 2) and the parameters $a=-9.3335$, $b=114.17$, $c=-337.56$, and $n=1.5465$ are obtained by fitting to the results of the numerical simulation. While the ejecta velocity $v_{\rm dyn}$ and the opening angle $\theta_{\rm dyn}$ of the ejecta are expressed as \citep{2017CQGra..34j5014D}: 
\begin{equation}
\begin{gathered}
v_{\mathrm{dyn}} \simeq\left[f_1\left(1+f_3 C_1\right) \frac{m_1}{m_2}+\frac{f_2}{2}\right]+(1 \leftrightarrow 2),  \\
\theta_{\mathrm{dyn}} \simeq \frac{-2^{\frac{4}{3}} v_\rho^2+2^{\frac{2}{3}}\left[v_\rho^2\left(3 v_z+\sqrt{9 v_z^2+4 v_\rho^2}\right)\right]^{\frac{2}{3}}}{\left[v_\rho^5\left(3 v_z+\sqrt{9 v_z^2+4 v_\rho^2}\right)\right]^{\frac{1}{3}}},
\label{theta:dyn}
\end{gathered}
\end{equation}
where $f_1=-0.3090$, $f_2=0.657$ $f_3=-1.879$, $v_{\rho}$ and $v_{\rm z}$ are the ejecta velocity component in the cylindrical coordinate given by \citet{2017CQGra..34j5014D}. {In the SMNS and stable NS case, the dynamical ejecta is set to  $m_{\rm dyn}\sim 0.01 M_{\odot}$ and the velocity $v_{\rm dyn}$ is set to $0.15c$ \citep{2019ApJ...880L..15M}. }

On the other hand, a fraction of NS decompressed matter may be centrifugally supported and therefore produce an accretion disk around the merger remnant, which may also contribute significantly to the ejecta mass through outflows driven by the neutrino wind near the symmetric axis \citep[e.g., ][]{2017LRR....20....3M}. The mass of this wind ejecta $m_{\rm wind}$ is often linked to the mass of the remnant disk $m_{\rm disk}$ by $\xi_{w}$, i.e., $m_{\rm wind}=\xi_{w}m_{\rm disk}$ \citep[e.g.,][]{2022ApJ...937...79C, 2023MNRAS.522..912Z}. In this work, we evaluate $m_{\rm disk}$ by the results of the numerical simulation \citep[e.g., ][]{2020PhRvD.101j3002K}:
\begin{equation}
m_{\rm disk}=m_1 \rm {\max}\left(5\times 10^{-4}, (aC_{1}+c)^d\right)
\label{mass:disk}
\end{equation}
where $a=-8.1324$, $c=1.4820$ and $d=1.7784$. The viscous torques of the accretion disks around the massive NSs or BH can also unbind ejecta matter, which may be much more massive than the wind ejecta driven by the neutrino wind on the viscous time scale. As in wind ejecta, we link the mass of the viscous ejecta $m_{\rm vis}$ with the mass of the remnant disk $m_{\rm disk}$ by $\xi_{v}$: $m_{\rm vis}=\xi_{v}m_{\rm disk}$ \citep[e.g.,][]{2022ApJ...937...79C}. In this work, we fix the fraction to be $\xi_{\rm w}=0.05$ and $\xi_{\rm v}=0.3$ for the component of wind and viscous ejecta \citep{2017arXiv171103982P,2022ApJ...937...79C}. 

Then the remnant mass $m_{\rm rem}$ of the BNS mergers can be estimated according to conservation of energy before the merger and after the remnant formed \citep[e.g., ][]{2022A&A...666A.174S}:
\begin{equation}
m_{\rm rem}=m_1+m_2-m_{\rm GW}-m_{\rm disk}-m_{\rm dyn},
\label{eq:mass}
\end{equation}
where $m_{\rm GW}$ is the GW energy radiated from the BNS merging process, which can be calculated using the fitting formula considering the inspiral to the post-merger stage \citep[e.g., ][]{Bernuzzi:2014kca, 2018PhRvL.120k1101Z}. In this work, we identify the remnant object as a BH (or short-lived HMNS) by its mass, i.e, if $m_{\rm rem}>1.2 M_{\rm TOV}$, while $1.0 M_{\rm TOV}<m_{\rm rem}<1.2 M_{\rm TOV}$ for the SMNS scenario and $m_{\rm rem}<1.0 M_{\rm TOV}$ for the stable NS scenario. For each mock BNS merger sample, we obtain a set $\mathbf{\Theta}=(z,m_1,m_2,r_1,r_2,\theta_{\rm v},m_{\rm dyn},m_{\rm vis},m_{\rm wind},v_{\rm dyn},m_{\rm rem})$ by the above equations~\eqref{mass:dyn}-\eqref{eq:mass}.

\begin{figure*}
\centering
\includegraphics[width=0.9\columnwidth]{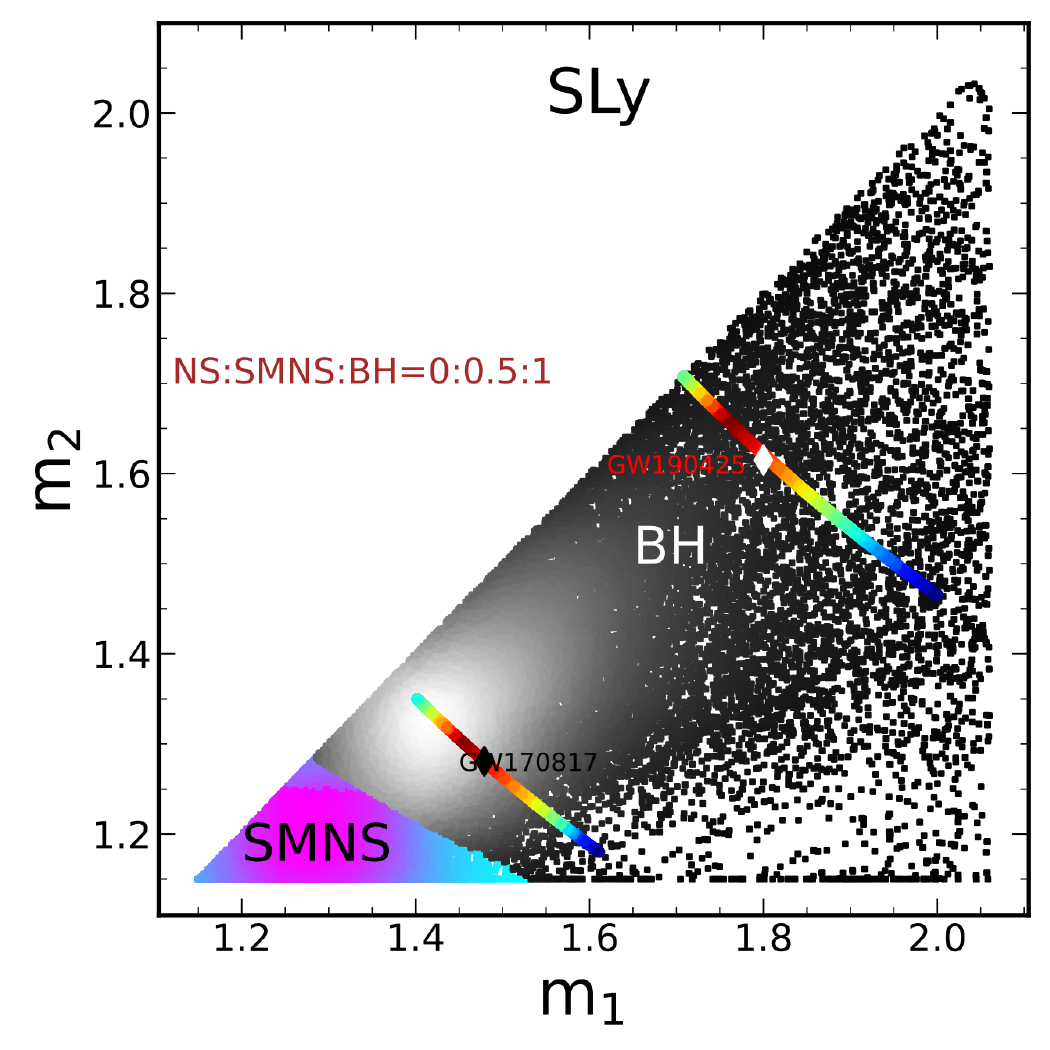}
\includegraphics[width=0.9\columnwidth]{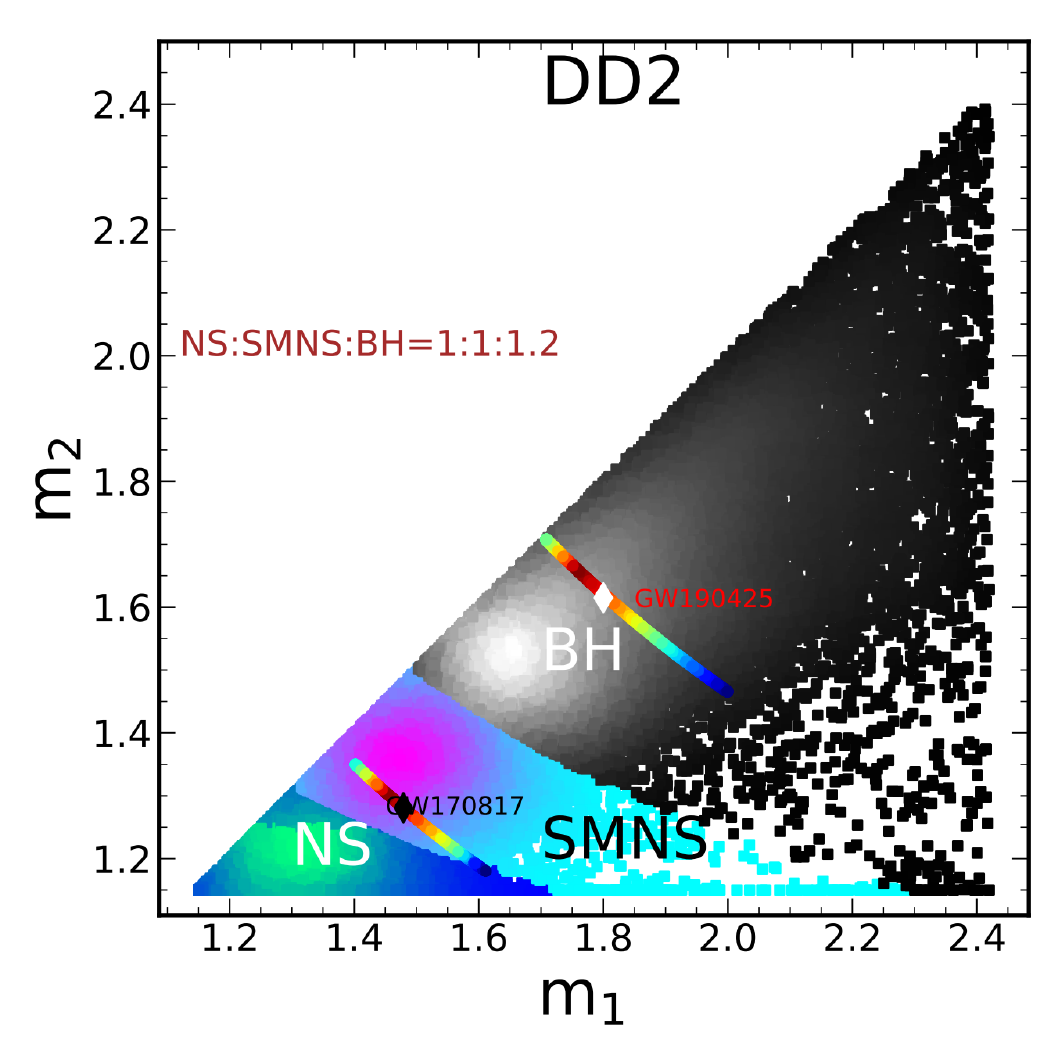}
\caption{
The distributions of the primary ($m_1$) and secondary ($m_2$) masses of BNS mergers are shown for different EOS, i.e., SLy (left panel) and DD2 (right panel). The colored region labeled with ``NS", ``SMNS" and ``BH" represent the mock BNS merger samples that result in a stable NS, a SMNS and a BH remnant, respectively. In both panels, brighter color regions indicate areas with a higher density of samples. The markers with color bars represent the median value and posterior distribution of GW170817 and GW190425, respectively. 
}
\label{fig:mass_dis}
\end{figure*}

Figure~\ref{fig:mass_dis} shows the mass spectrum of our mock BNS samples assuming either SLy or DD2 EOS. The number ratio of stable-NS/SMNS/BH remnants is $\sim 0:0.5:1$ if adopting the SLy EOS (left panel), i.e., no stable-NS remnant is produced, while it is $\sim 1:1:1.2$ if adopting the DD2 EOS (right panel). The different number ratio resulting from different EOSs is mainly due to different $M_{\rm TOV}$ and different mass ejecta properties for different EOSs. We also mark the median value and the posterior distribution of GW170817 in the $m_1-m_2$ plane obtained by fitting the GW signals to the  {IMRPhenomPv2NRT-lowSpin} GW model \citep[e.g., ][]{2017PhRvL.119p1101A}. According to the constraints obtained by \citet{2017ApJ...850L..19M}, using multi-messenger observations, the SLy EOS matches much better than the stiffer EOS, for example, DD2. Therefore, we come to the conclusion that the remnant of GW170817 is more likely to be a BH/HMNS, rather than an magnetar, which is consistent with the constraint from the tidal deformabilities \citep[e.g., ][]{2017PhRvL.119p1101A, 2017ApJ...850L..19M}. Besides, we also plot the results of GW190425 by \citet{2020ApJ...892L...3A} in the Figure and apparently the merger remnant of GW190425 is likely to be a BH. 

In principle, both the stable-NS and SMNS can act as extra energy reservoirs that enhance merger-nova signals by injecting spin-down energy via strong magnetic wind and their difference lies within their lifetime $\tau$. Since the peak of the merger-nova signals typically happens at  $t_{\rm d}=1$ after the formation of remnants, we focus on the merger-nova signals with energy injection lasting for at least 1 day and thus further classify the remnants into two categories by their types at $t_{\rm d}=1$ day, i.e.,  survived magnetar and BH (hereafter we omit the explicit specification of $t_{\rm d}=1$ day for simplicity). By constructing the relationship between $\tau$ and the mass of the remnant $m_{\rm rem}$ from the results of both numerical simulations and observations on GW170817 and X-ray plateaus of sGRB signals, we find that: if the EOS is SLy, the magnetar is low-mass SMNS with $\lesssim 2.20 M_{\odot}$. If the EOS is DD2, the magnetar is stable-NS with mass $\lesssim 2.42 M_{\odot}$. A more detailed description is presented in the Appendix ~\ref{app:sd}. With the two categories of mock BNS samples, we calculate the merger-nova luminosity function by introducing two different merger-nova emission models from the parameter sets $\mathbf{{\Theta}}$. 

\subsection{Merger-nova signal}
\label{sec:model}

Generally speaking, the merger-nova is driven by the radioactive decay of heavy elements, such as lanthanide elements, produced by the rapid neutron capture process (r-process) in the mass ejecta of BNS mergers. Due to the different formation mechanisms of the mass ejecta discussed above, their temperature and opacity may be significantly different. In this paper, we adopt the anisotropic multi-component model as the merger-nova model for the case with BH remnants, which is in general red at equatorial direction and blue at polar direction. More detailed description can be found in Appendix ~\ref{app:model}. 
We then obtain the best-fit of the 9 model parameters, including energy normalization $\epsilon_{0}$, lanthanide-rich flat temperature $T_{f}^{\rm LA}$, lanthanide-free flat temperature $T_{f}^{\rm Ni}$, low-elevation opacity $\kappa_{\rm low}$, high-elevation opacity $\kappa_{\rm high}$, wind ejecta opacity $\kappa_{\rm wind}$, viscous ejecta opacity $\kappa_{\rm vis}$, RMS velocity of viscous ejecta $v_{\rm rms}^{\rm vis}$ and RMS velocity of wind ejecta $v_{\rm rms}^{\rm wind}$, by using the $u$, $g$, $r$, $i$, and $z$ band LCs of AT2017gfo with the Bayesian approach, fixing the luminosity distance estimated by GW170817, i.e, $d_{\rm L}=40 \rm Mpc$, and the mass of the ejecta components from numerical simulation results. For the {magnetar} remnant scenario, we introduce an additional isotropic magnetic wind transfer fraction $\xi_{\rm B}$ from the spindown energy of the {magnetar} $L_{\rm sd}$ to the merger-nova luminosity and fit the value of $\xi_{\rm B}$ by the extreme luminous merger-nova associated with the short GRB 130603B \citep{2013Natur.500..547T}. More detailed information on the merger-nova model and fitting procedure can be found in the Appendix~\ref{app:fit}. 

By assuming the best-fit parameters to be generic for all merger-nova signals and varying the other parameters according to our mock BNS merger samples generated in Section~\ref{sec:sample}, we can estimate the luminosity function $\frac{d^2\dot{\Phi}_{\nu}({ M_{\rm AB}})}{dM_{\rm AB}dV_{\rm c}}$ at frequency $\nu$ with respect to the absolute magnitude $M_{\rm AB}$, where $\dot{\Phi}_{\nu}({ M_{\rm AB}})$ can be calculated as
\begin{equation}
\dot{\Phi}_{\nu}(M_{\rm AB}>M(F_{\nu}))=\int d\mathbf{\Theta} \dot{N}(\mathbf{\Theta})G_{\nu}(\mathbf{\Theta}),
\label{eq:lumi}
\end{equation}
with:
\begin{equation}
\begin{aligned}
G_{\nu}(\mathbf{\Theta})=& H(F(\mathbf{\Theta})_{\rm magnetar,\nu}-F_{\nu}) \\
    & \times H(1.2M_{\rm TOV}-m_{\rm rem}(\mathbf{\Theta}))H( m_{\rm rem}(\mathbf{\Theta})-M_{\rm TOV})\\
    &+H(F(\mathbf{\Theta})_{\rm BH,\nu}-F_{\nu}) H( m_{\rm rem}(\mathbf{\Theta})-1.2M_{\rm TOV}), 
\end{aligned}
\end{equation}
where $H$ is the Heaviside function, $F(\mathbf{\Theta})_{\rm BH,\nu}$ and $F(\mathbf{\Theta})_{\rm magnetar,\nu}$ are the flux observed at frequency $\nu$ calculated by BH and magnetar merger-nova models for given mock BNS merger sample parameters $\mathbf{{\Theta}}$, respectively. By the above equation, we estimate the luminosity function in the $u$, $g$, $r$, $i$, and $z$ bands with both EOSs of the BNS mergers, i.e., SLy and DD2. By the above expressions, we estimate the luminosity function $\dot{\Phi}_{\nu}(M_{\rm AB})$ in different frequency bands of the merger-nova signal emitted by our mock BNS merger samples for both EOSs, i.e., SLy and DD2.  

\subsection{Detection efficiency}
\label{subsec:feff}

The most probable strategy for merger-nova searching is the Target of Opportunity (ToO) strategy, by which one uses telescopes to scan over the possible sky-area constrained by GW signals. The detection probability $P$ of ToO for a merger-nova signal with a given allocated observation time $t_{\rm obs}$ (assumed to be 1 hour in this paper) can be defined as
\begin{equation}
\begin{aligned}
      P={\rm min}
      \bigg(1,\frac{\Omega_{\rm{ FOV}}}{\Delta\Omega}\frac{t_{\rm obs}-t_{\rm exp}}{t_{\rm exp}},\frac{\Omega_{\rm{ FOV}}}{\Delta\Omega}\frac{\Delta T-t_{\rm exp}}{t_{\rm exp}}\bigg),
\label{eq:feff}
\end{aligned}
\end{equation}
where $\Omega_{\rm FOV}$ is the field of view (FOV), $t_{\rm exp}$ is the exposure time required to reach a given limiting magnitude, $\Delta T$ is the time interval of the light curve brighter than the limiting magnitude of the telescope $m_{\rm lim}$, and $\Delta\Omega$ represents the localization area constrained by GW observation. {Note that in the above expression we have assumed that two observations are required to monitor the luminosity change of the merger-nova signals.} If $\Delta T =0 $, i.e., the merger-nova is always fainter than the limiting magnitude, the detection probability is $P=0$. For the merger-novae associated with BNS mergers with any given set of properties, e.g., within the same redshift bin or mass bin, the average detection efficiency should be 
\begin{equation}
\langle f_{\rm eff} \rangle = \frac{1}{N_{\rm GW}} \sum_{i=1}^{N_{\rm GW}} P_{i},
\end{equation}
where $N_{\rm GW}$ is the total number of GW-detected BNS mergers with that set of properties, and $P_{i}$ is the probability of detection of the $i$-th BNS merger. For demonstration, we adopt the Chinese Space Station Telescope (CSST), with an FOV of $\Omega_{\rm FOV}=1.1\rm deg^2$ \citep{2019ApJ...883..203G}, to calculate the average detection efficiency $\langle f_{\rm eff} \rangle$ of merger-nova signals. We choose the CSST filter $u$, $r$, and $z$, as an example, with the corresponding limiting magnitudes $m_{{\rm lim}}$ of $25.4$, $26.0$, and $25.2$\,mag, and the exposure time of a single pointing is set as $300$\,s.

\subsection{GW detection and localization}
\label{subsec:GW}

\begin{figure*}
\centering
\includegraphics[width=0.9\columnwidth]{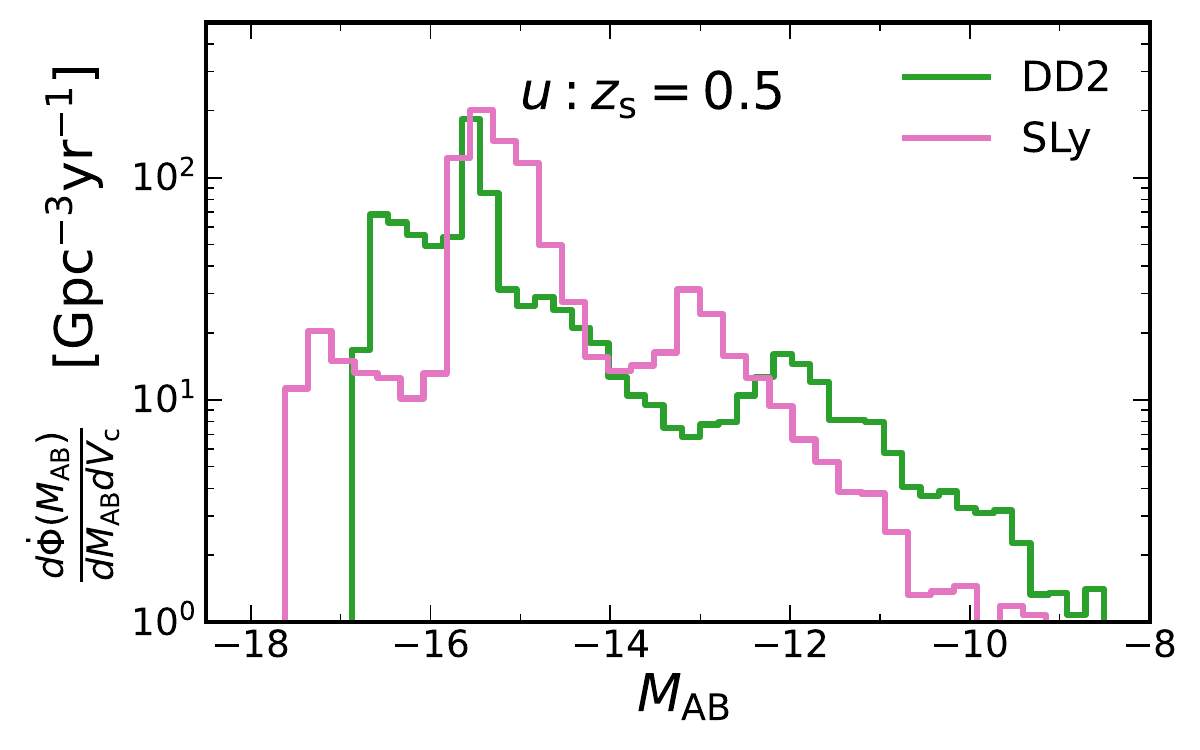}
\includegraphics[width=0.9\columnwidth]{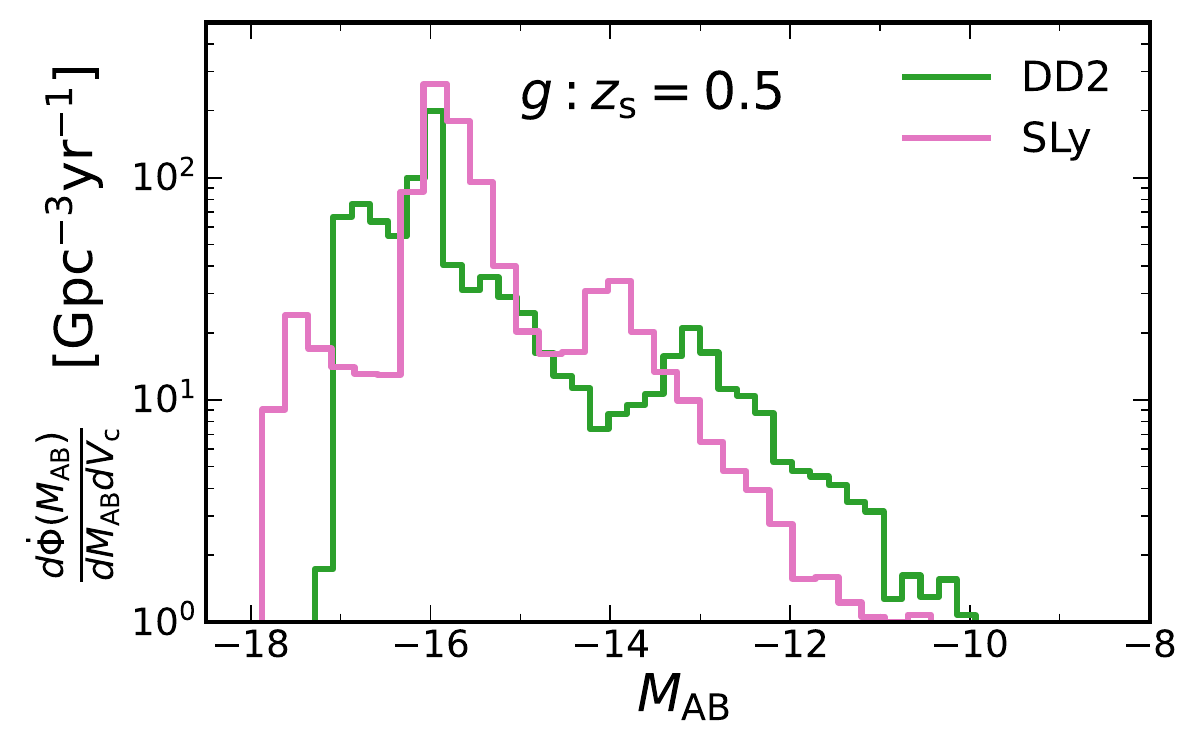}
\includegraphics[width=0.9\columnwidth]{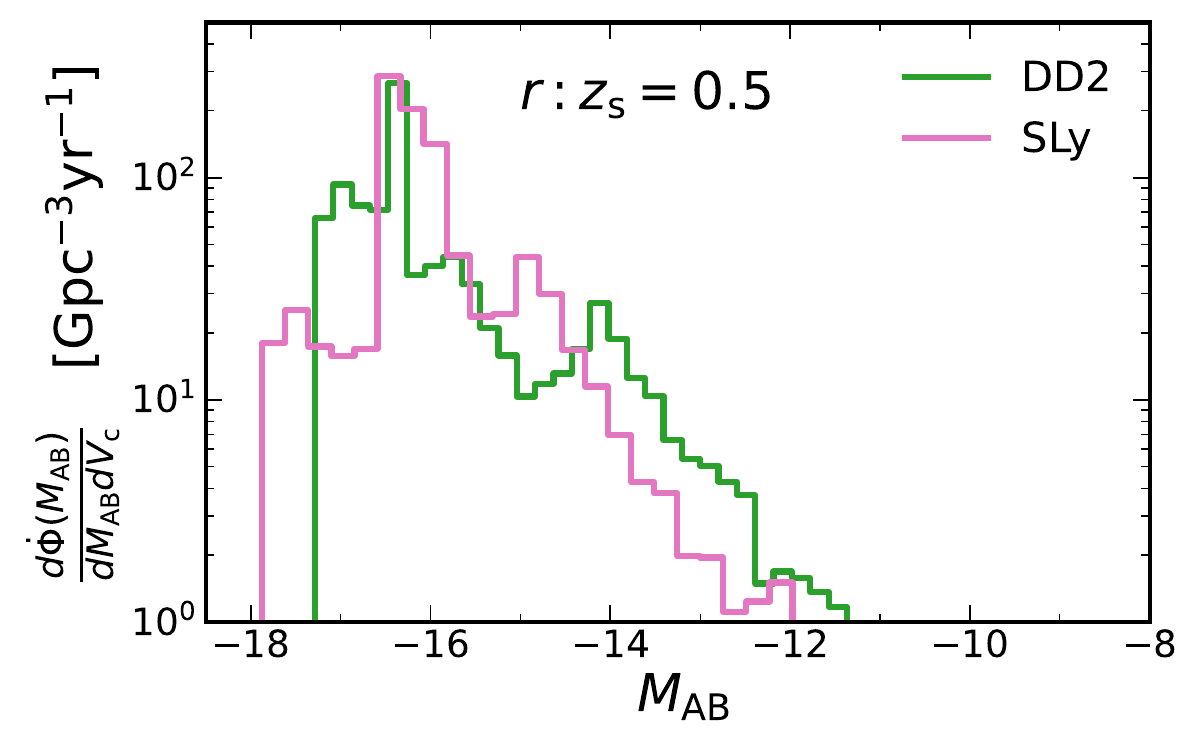}
\includegraphics[width=0.9\columnwidth]{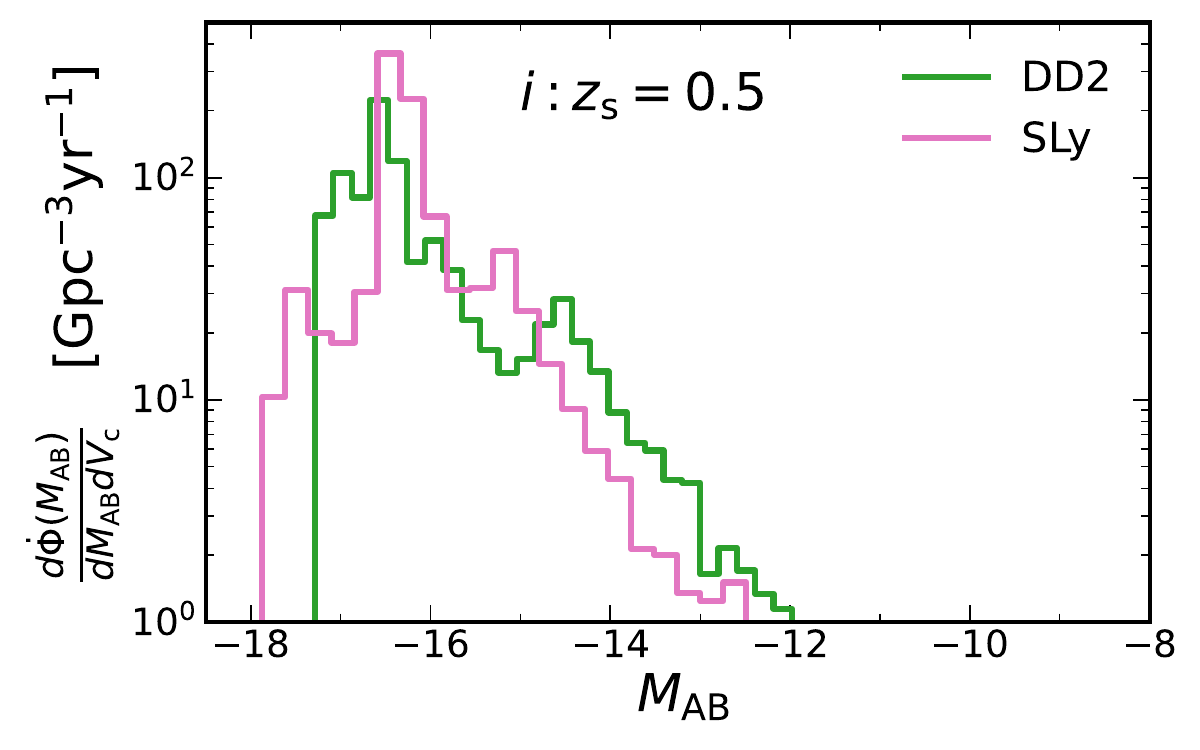}
\includegraphics[width=0.9\columnwidth]{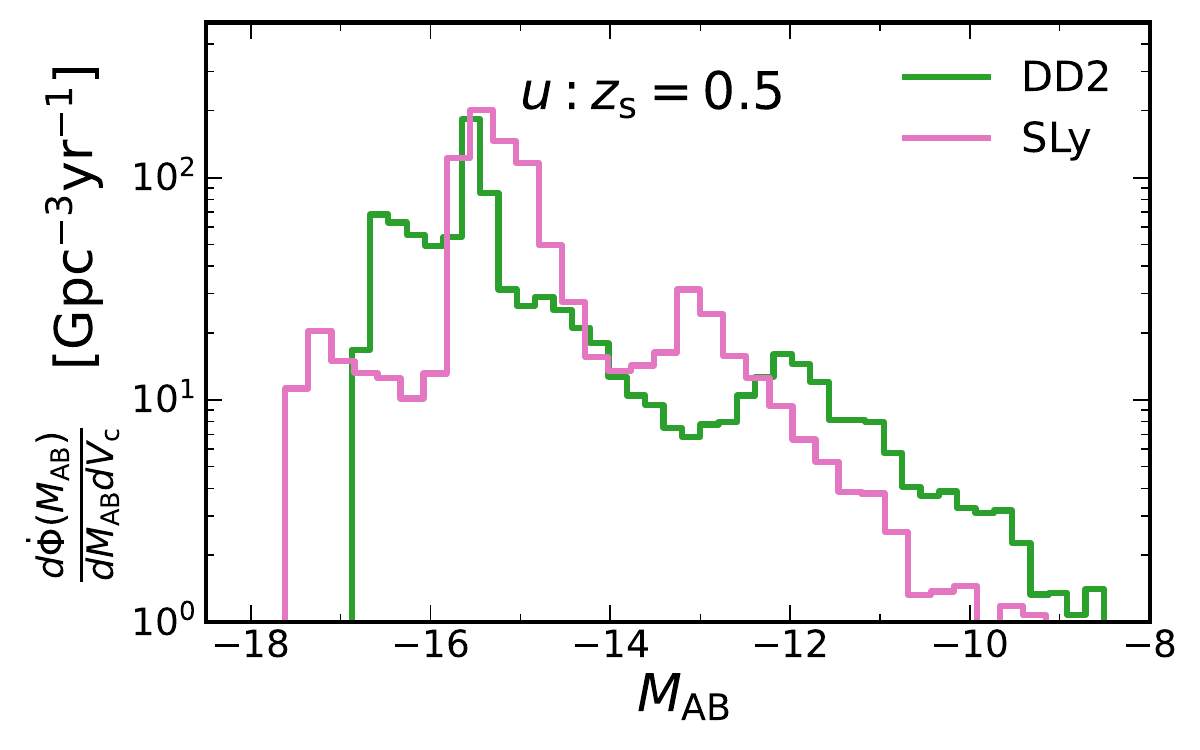}
\caption{
{The luminosity function ${d\dot{\Phi}(M_{\rm AB})}/{dM_{\rm AB}dV_{\rm c}}$ of the merger-nova signals at $z_{\rm s}=0.5$ produced by BNS mergers with different EOS, specifically DD2 (green lines) and SLy (pink lines), defined by the intrinsic absolute magnitude at $t_{\rm d}=1$\,day in different bands, i.e., $u$, $g$, $r$, $i$ and $z$ bands.  }
}
\label{fig:lf}
\end{figure*}

For GW detection, we calculate the signal-to-noise ratio $\rho_{\rm GW}$ and the localization precision $\Delta\Omega$ using the following Monte Carlo procedure. We first assign the orientation angles to each BNS merger event, i.e., ( $\theta_{\rm s}$, $\phi_{\rm s}$, $i$, $\psi$), which are all uniformly and randomly sampled in the sky. {Here $\theta_{\rm s}$ and $\phi_{\rm s}$ are the polar and azimuthal angle in the sky}, while $i$ and $\psi$ give the orientation of the source with respect to the detector. Then we adopt the standard package  {pyCBC} \citep[][]{2019PASP..131b4503B} to generate the GW waveform for each BNS merger, adopting the phenomenological model  {IMRPhenomPv2-NRTidalv2}, of which the total strain $h(f)$ received by a GW detector is
\begin{equation}
h(f)=F_{+}(f)h_{+}(f)+F_{\times}(f)h_{\times}(f),
\end{equation}
where $F_{+}$ and $F_{\times}$ are the detector's pattern functions for the $+$ and $\times$ polarization, of which the explicit expressions in the time domain (i.e., $\Tilde{F}_{+,\times}(t)$)  are periodic functions of time with a period equal to one sidereal day, due to the diurnal motion of the Earth \citep[e.g.,][]{PhysRevD.58.063001,PhysRevD.81.062003,2018PhRvD..97f4031Z}.

Here we define the whitened GW data sets of a GW network composed of $n$ detectors (e.g., $n=1$ for a single detector and $n=2$ for two detectors) as
\begin{equation}
\hat{\mathbf{d}}(f)=\left(\frac{A_1(f)h_1(f)}{\sqrt{S_{1}(f)}},\frac{A_2(f)h_2(f)}{\sqrt{S_{2}(f)}}, ..., \frac{A_{n}(f)h_{n}(f)}{\sqrt{S_{n}(f)}}\right),
\end{equation}
where $A_{n}=e^{-2\pi i f((\hat{r}_{n}-\hat{r}_{1})\cdot \hat{n}_{\rm GW})}$ is the phase transfer function, $\hat{r}_{n}$ is the location vector of the $n$-th detector, $\hat{n}_{\rm GW}$ is the unit direction vector of the GW source, and $S_{n}$ denotes the one-sided power spectrum of the corresponding $n$-th GW detector.  Then the optimal squared SNR is given by
\begin{equation}
\varrho_{GW}^2=\left\langle\hat{\mathbf{d}}(f)|\hat{\mathbf{d}}(f)\right\rangle,
\label{eq:SNR}
\end{equation}
where the angular bracket denotes an inner product. For any two vector functions $\hat{\mathbf{a}}(f)$ and $\hat{\mathbf{b}}(f)$, this inner product is defined as
\begin{equation}
\langle \hat{\mathbf{a}}(f)| \hat{\mathbf{b}}(f)\rangle=2\sum_{j} \int_{f_{\text {min }}}^{f_{\text {max }}}\left\{a_j(f) b_j^{*}(f)+a_j^{*}(f) b_j(f)\right\} d f ,
\end{equation}
where $j$ denotes the $j$-th component of the vector, ${f_{\rm min}}$ and ${f_{\rm max}}$ are the lower and upper frequency limits of the GW waveforms.

The localization error for each BNS GW source may be estimated according to the Fisher information matrix \citep[e.g.,][]{PhysRevD.58.063001, 2018PhRvD..97f4031Z, 2022ApJ...940...17C, 2023MNRAS.522.2951Z}, $\Gamma_{jk}$, defined as
\begin{equation}
\Gamma_{j k}=\left\langle\partial_{j} {\hat{\mathbf{d}}(f)}\mid \partial_{k}\hat{\mathbf{d}}(f)\right\rangle,
\end{equation}
where $\partial_j$ and $\partial_k$ denote the partial derivative with respect to the $j$-th and $k$-th parameter, respectively. Once the Fisher matrix is determined, the covariance matrix of the location of a GW source in the celestial coordinates is given by
\begin{equation}
\rm Cov(\theta_{\rm s},\phi_{\rm s})=\Gamma^{-1}.
\label{eq:error}
\end{equation}
With this total covariance matrix, we get the localization errors for a BNS merger in solid angle as
\begin{equation}
\Delta \Omega_{\rm GW}=2 \pi\left|\sin \theta_{\rm s}\right| \sqrt{\left\langle\Delta \theta_{\rm s}^{2}\right\rangle\left\langle\Delta \phi_{\rm s}^{2}\right\rangle-\left\langle\Delta \theta_{\rm s} \Delta \phi_{\rm s}\right\rangle^{2}},
\label{eq:omega}
\end{equation}
where $\Delta \theta_{\rm s}$ and $\Delta \phi_{\rm s}$ is the standard deviation obtained from the covariance matrix. 

\section{Results}
\label{sec:results}

\begin{figure*}
\centering
\includegraphics[width=0.75\columnwidth]{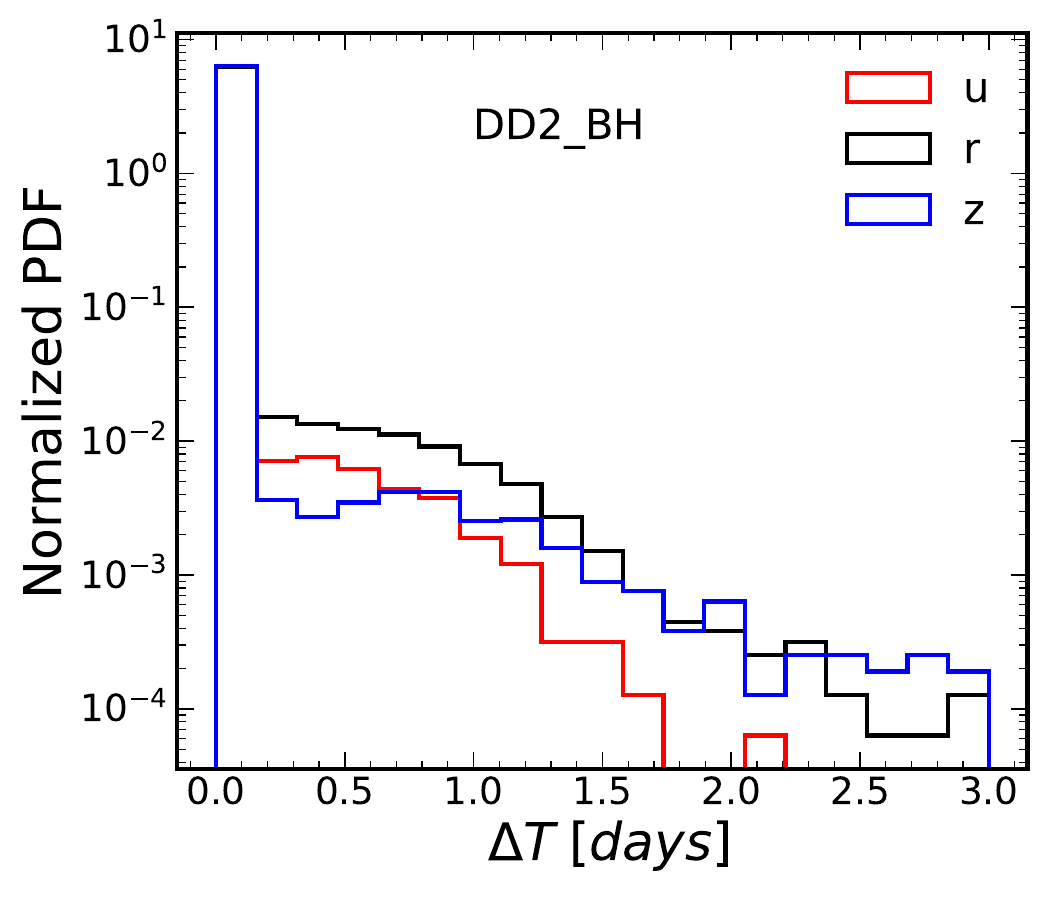}
\includegraphics[width=0.75\columnwidth]{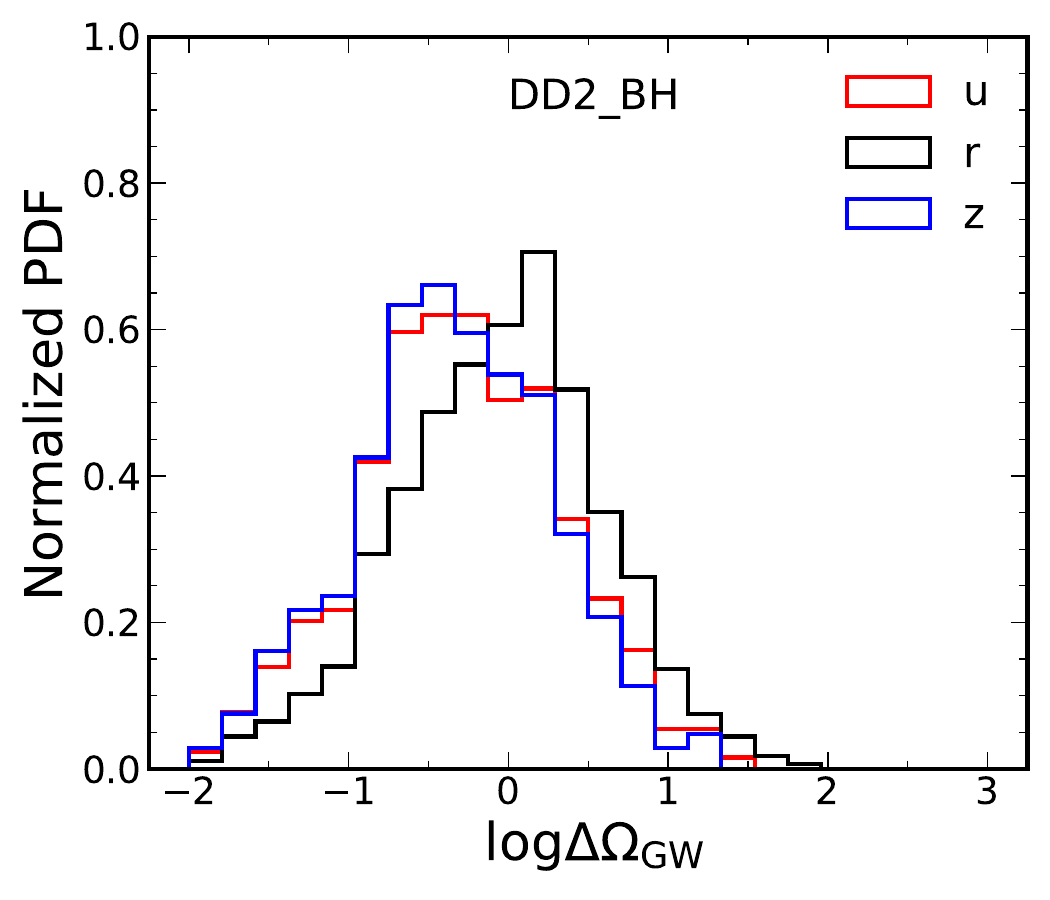}
\caption{
The normalized probability distribution of BNS mergers with BH remnant associated with merger-nova signals, assuming that the EOS is DD2. {The left panel shows the results for the time span above the limiting magnitudes of CSST (if always below the detection threshold, $\Delta T=0$). The right panel shows the results for the localization precision of GW signals with Cosmic Explorer (CE) associated with merger-nova signals with $\Delta T>0$. The red, black, and blue histograms represent the results for the $u$, $r$, and $z$ filters of CSST. }
}
\label{fig:dd2_BH}
\end{figure*}

\begin{figure*}
\centering
\includegraphics[width=0.75\columnwidth]{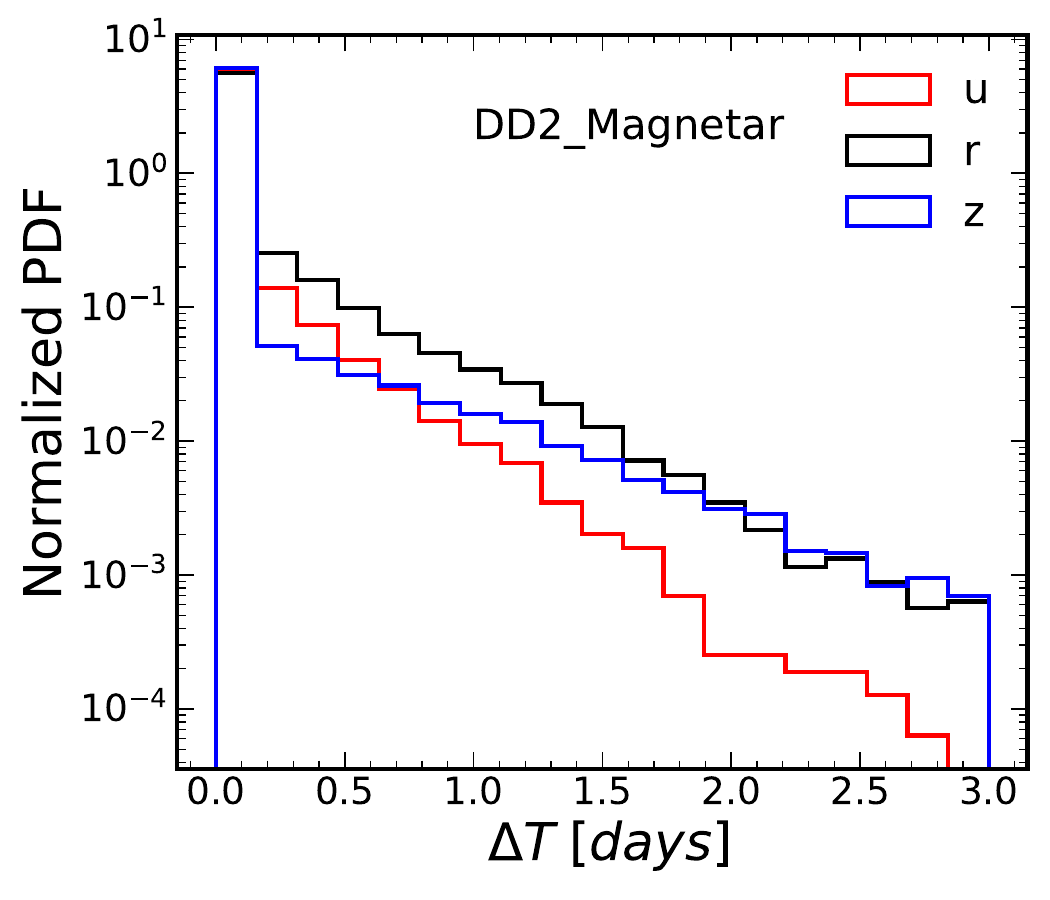}
\includegraphics[width=0.75\columnwidth]{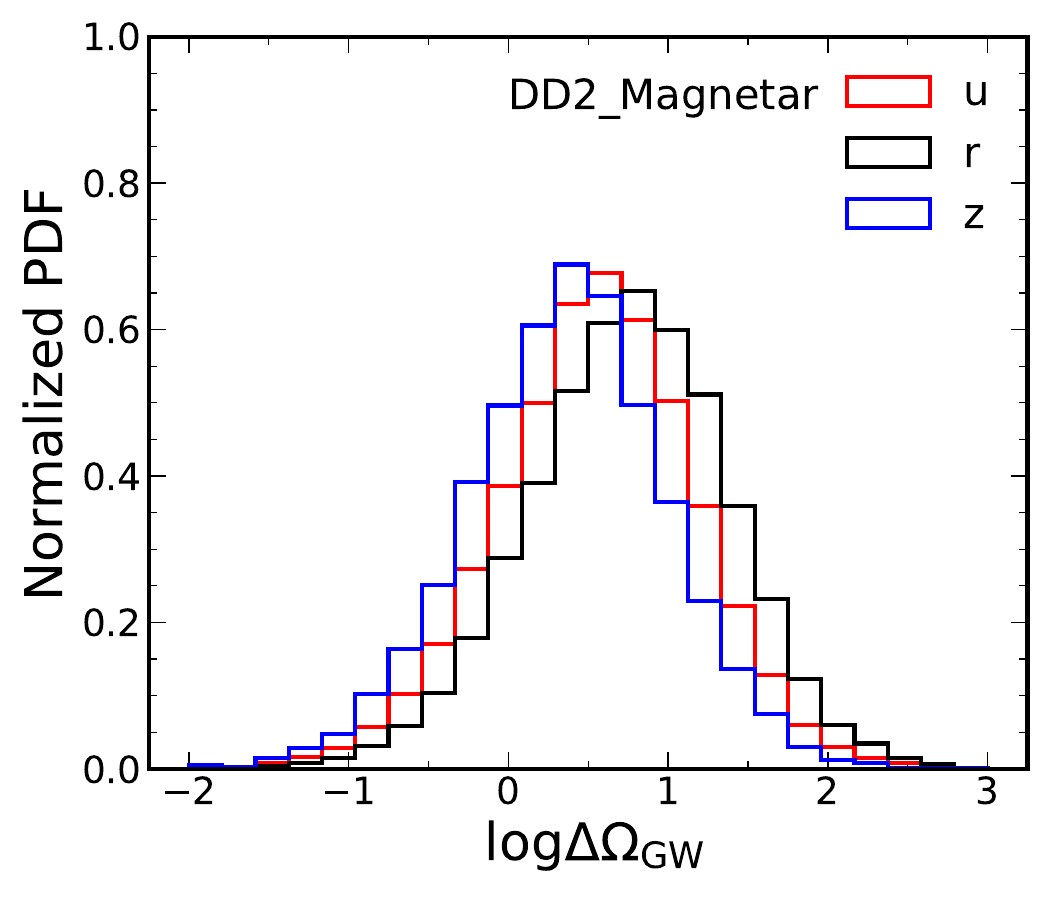}
\caption{
The legend is the same as in Figure~\ref{fig:dd2_BH}, except that the merger remnant is a {magnetar}.
}
\label{fig:dd2_smns}
\end{figure*}

\begin{figure*}
\centering
\includegraphics[width=0.75\columnwidth]{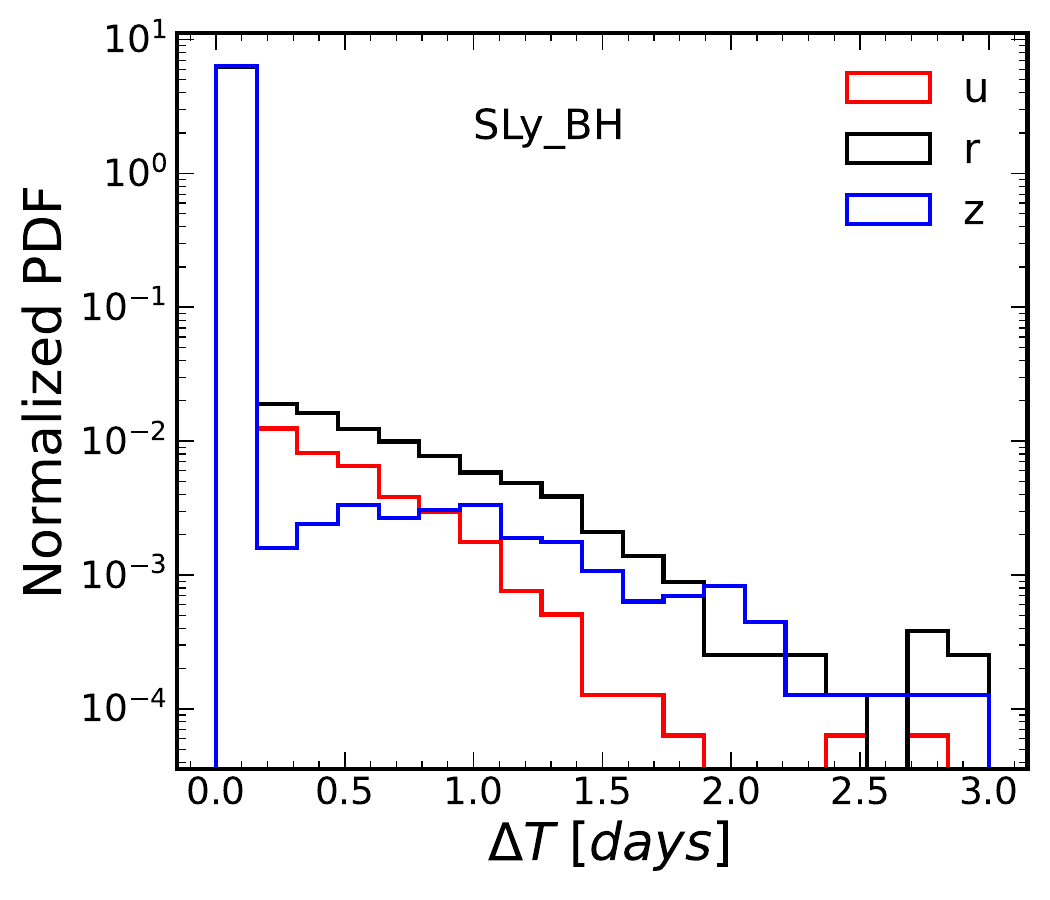}
\includegraphics[width=0.75\columnwidth]{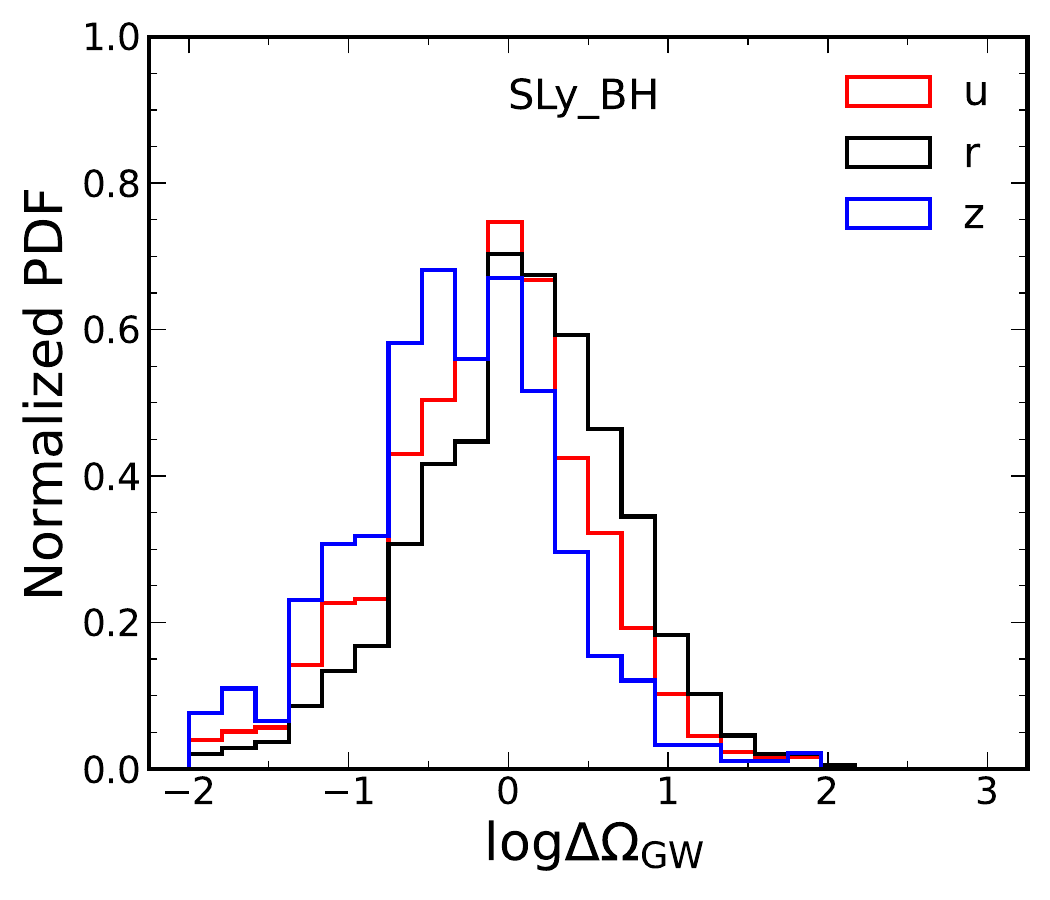}
\caption{
The legend is the same as in Figure~\ref{fig:dd2_BH}, except that the EOS is assumed to be SLy.
}
\label{fig:sly_BH}
\end{figure*}

\subsection{Luminosity Function}

Figure~\ref{fig:lf} shows the intrinsic luminosity function ${d\dot{\Phi}(M_{\rm AB})}/{dM_{\rm AB}dV_{\rm c}}$ of the merger-nova signals in different bands at $t_{\rm d}=1$\,day produced by the BNS mergers with different EOSs at $z_{\rm s}=0.5$, i.e., SLy (pink) and DD2 (green). Note that the overall distributions at different redshifts are qualitatively similar to each other, although the exact values may differ. 

As seen in the figure, the most significant features of the luminosity functions from both EOSs are the three peaks at different absolute magnitudes for all bands. The first peak, around $\sim -17.5$\,mag, represents BNS mergers with survived magnetar remnants at $t_{\rm d}=1$ day. This peak is due to the substantial enhancement in the radiation caused by the isotropic magnetic wind driven by the spin-down energy of the central magnetar. The width of this peak is mainly influenced by the choice of the energy transformation factor $\xi_{\rm B} $ of $L_{\rm sd}$, which is assumed to be uniformly distributed within the range of $[0,0.2]$. (see Appendix B for more details). Additionally, the scatter of this peak in different bands is relatively small, mainly due to the frequency-independent contribution of $L_{\rm sd}$. The total intrinsic event occurrence rate of the magnetar merger-nova signals for SLy is significantly smaller than that for DD2, which is mainly attributed to the difference in the number of BNS merger remnants with masses in the survived magnetar mass range, directly related to the assumed EOS. For example, the occurrence rate of merger-nova signals with magnetar remnants at redshift interval $z_{\rm s}\in [0.4,0.6]$ are about $\sim 2.50\times 10^3~\rm yr^{-1}$ and $\sim 1.05\times 10^4~ \rm yr^{-1}$ for SLy and DD2 respectively.

The second and third peaks are natural results of the ANI-DVN model adopted in this paper to describe the LCs of merger-nova signals in the BH scenario, mainly due to the range of possible viewing angles within $[0,\pi]$. We refer the reader to \citet{2023MNRAS.522..912Z}, who defined the luminosity function of the kilonovae (here referred to as the merger-novae in the BH scenario) by their peak luminosity. Unlike the first peak, the locations of the latter two peaks vary significantly with frequency. For example, the second peak occurs around $-15.0$, $-15.5$, $-16.0$, $-16.1$, and $-16.3$\,mag in the $u$, $g$, $r$, $i$, and $z$ bands, respectively, for the SLy EOS. It is evident that the redder the band, the brighter the merger-nova signals. This is a direct result of the fitted parameters in the LC models obtained from the observation of AT2017gfo, i.e., it is rather brighter in redder bands at $t_{\rm d} \sim 1$\,day. The occurrence rate of merger-nova signals with BH remnants at redshift interval $z_{\rm s}\in [0.4,0.6]$ are about $\sim 2.95\times 10^4~\rm yr^{-1}$ and $\sim 2.13\times 10^{4} ~\rm yr^{-1}$ . In conclusion, the luminosity function of the BNS merger-nova at $t_{\rm d}=1$\,day reveals that the relative positions of the first and the latter two peaks are influenced by the injection of the spin-down energy of the {magnetar} remnant. The more efficient the energy ejection by the magnetic wind, the brighter the first peak. The relative heights of the three peaks provide insight into the EOS of the BNS. The stiffer the EOS results in a higher first peak compared to the latter two peaks. Therefore, the maximum mass $M_{\rm TOV}$ may be robustly constrained using the luminosity function of their merger-nova signals, though these signals are strongly dependent on the radiation model.

\begin{figure*}
\centering
\includegraphics[width=0.75\columnwidth]{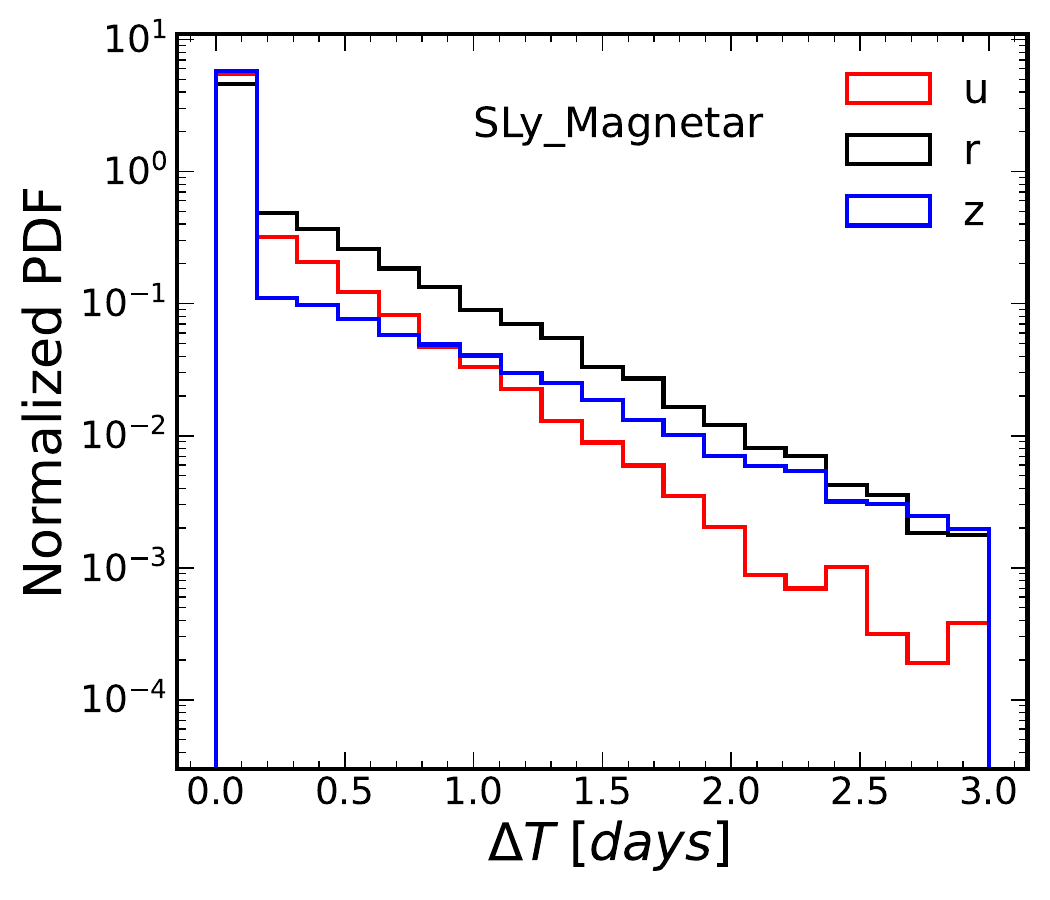}
\includegraphics[width=0.75\columnwidth]{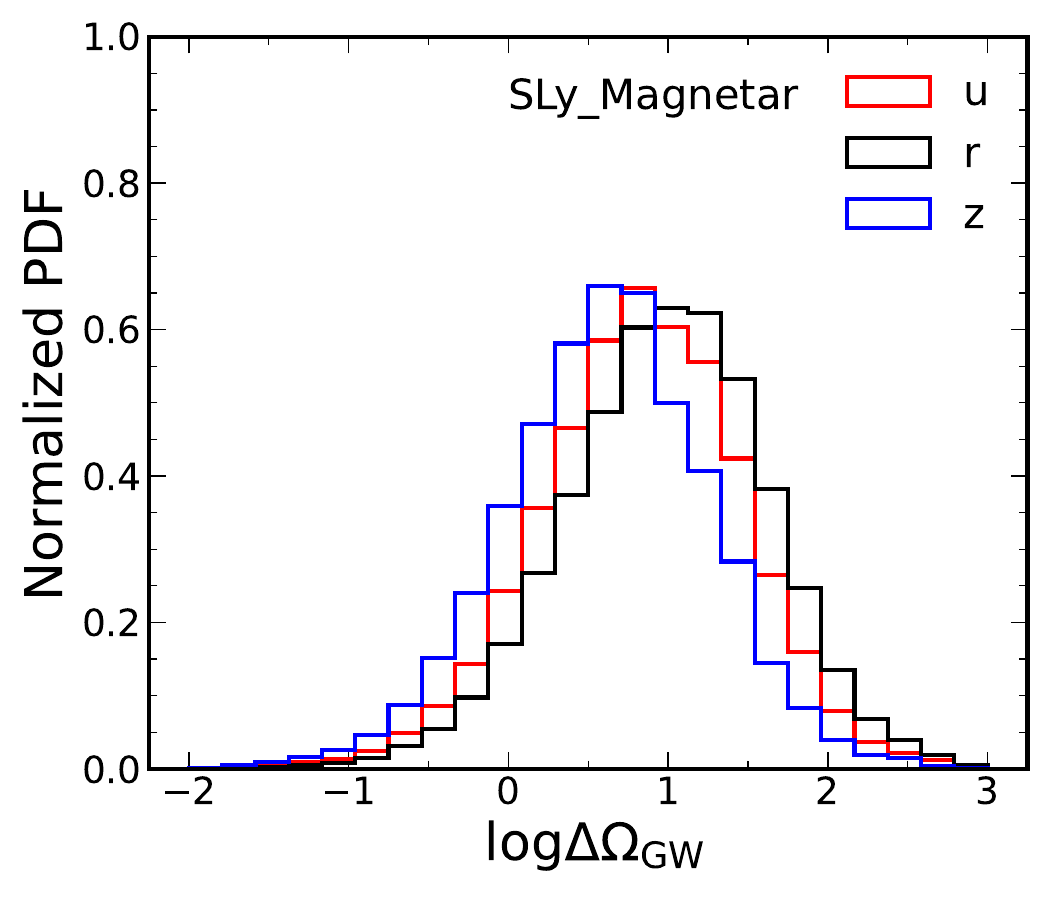}
\caption{
The legend is the same as in Figure~\ref{fig:dd2_BH}, except that the EOS is assumed to be SLy and the merger remnant is a {magnetar}.
}
\label{fig:sly_smns}
\end{figure*}

Before entering the presentation of the ToO detection efficiency of the merger-nova signals, we would like to further discuss the possible uncertainties to construct such a luminosity function with various remnants from the observational aspect. It can be seen that the determination of the apparent magnitude of a given merger-nova may suffer from several observational effects, including the sampling of light curves and extinction. According to the observations of merger-nova signals in literature,  i.e., AT2017gfo associated with GW170817 \citep[e.g.,][]{2017ApJ...848L..12A} and that associated with GRB230307A detected by JWST \citep[e.g.,][]{2024Natur.626..737L}, the typical uncertainties of the apparent magnitudes at $t_{\rm d}=1$\,day (the counting time adopted in this paper) determined from observations is $\lesssim 0.1$ mag. In the context of this work, the magnitude differences between the first and second peaks of the predicted merger-nova luminosity functions are about $\sim 2$ magnitudes. Therefore, we conclude that the uncertainty in the determination of the apparent magnitude at the time of $1$\,day after the merger may only slightly affect the classification of the first and second peaks in the luminosity functions. 

{We also note that the distinguish of EOS by the shape of merger-nova luminosity function may suffer from  the measurement uncertainties and counting noise when the detection number is not sufficient. Here we perform a simple Kolmogorov-Smirnov (KS) test based on Monte-Carlo simulations to estimate the minimum number of detection required for confident identification, which may be instructive for further investigation. First, we randomly draw different numbers of samples from the LC templates constructed for both EOSs and assign Gaussian uncertainties to the absolute magnitude with typical dispersion of $\sigma\sim 0.1$ mag to model real detection. Then, we estimate the KS test power $1-\beta_{\rm KS}$ (i.e., the statistical power to detect a difference between two distributions), by assuming two different Type-I error hypothesis testing significance level $\alpha_{\rm KS}=0.05$ and $\alpha_{\rm KS}=0.10$, which are the typical choice for exploratory and confirmatory studies respectively \citep[e.g.,][]{myors2023statistical}. With the average of $10^{4}$ random realizations, our Monte-Carlo simulation reveals that the minimum detection number required for confidence level $\alpha_{\rm KS}=(0.05,0.10)$ are about $\sim (55,75)$ and $\sim (25,40)$, when adopting the benchmark high ($1-\beta_{\rm KS}=0.8$) and moderate ($1-\beta_{\rm KS}=0.5$) Cohen standards, respectively \citep{cohen1988statistical}. Future powerful space-borne telescopes have the capability to detect a factor of several times of this estimated minimum number of merger-nova signals with years of accumulation due to their deep limiting magnitude \citep[e.g.,][]{2019MNRAS.485.4260S,2022ApJS..258....5A,2025RAA....25c5018Z}. In summary, we are optimistic about the construction of their luminosity function via the observation of merger-nova signals in the upcoming next generation GW detection era.}
 
\subsection{$\Delta T$ and $\Delta \Omega_{\rm GW}$}

As discussed in Section~\ref{subsec:feff}, the detection efficiency of merger-nova signals associated with GW-detected BNS mergers depends on both the localization area of the GW signal and the time interval $\Delta T$ within which the magnitude is brighter than the limiting magnitude $m_{\rm lim}$. It is important to note that the exact detection rate is strongly dependent on the observation strategy, specifically the time allocated to each BNS merger. Therefore, we do not provide an exact number but instead focus on the normalized distribution of these parameters for BNS mergers associated with observable merger-nova signals.  

Figure~\ref{fig:dd2_BH} shows the normalized probability distribution of BNS mergers with BH remnant associated with merger-nova signalS, assuming that the EOS is DD2. In this figure, the limiting magnitude of CSST for  $u$, $r$ and $z$ bands are set to be $25.4$, $26.0$, and $25.2$ mag respectively. In the left and right panels, we show the resultant distribution of the time span $\Delta T$ and the localization precision $\Delta \Omega$ while the yellow, green and blue color represents the results adopting the $u$, $r$, and $z$ filters of CSST, from which we can reach the following conclusions. 
{First, only a small fraction of $\sim 0.63/1.42/0.52\%$ of merger-nova signals can have $\Delta T>0$ in $u$, $r$, and $z$ bands of CSST respectively, due to their different limiting magnitudes. In addition, one may see that both the results in the $r$ and $z$ bands are much more extended than in the $u$ band. This can be partly explained by the merger-nova model adopted, producing systematically brighter LC at redder bands calibrated by the observation of AT2017gfo. } Second, the localization precision of BNS mergers with observable merger-nova signals is likely to be a logarithmic Gaussian distribution for all three filters. Their median value and standard deviation are similar, i.e., $\mu\sim 0.49$, $0.94$, $0.43$\,deg$^2$ and $\sigma\sim 0.67$, $0.66$, and $0.66$\,dex, respectively. This localization precision hints that about half of merger-nova signals beyond the limiting magnitude can be efficiently searched with a single CSST scan, benefiting from the large FOV ($\Omega_{\rm FOV}=1.1\rm deg^2$) of CSST \citep{2019ApJ...883..203G}. 

\begin{figure*}
\centering
\includegraphics[width=1.2\columnwidth]{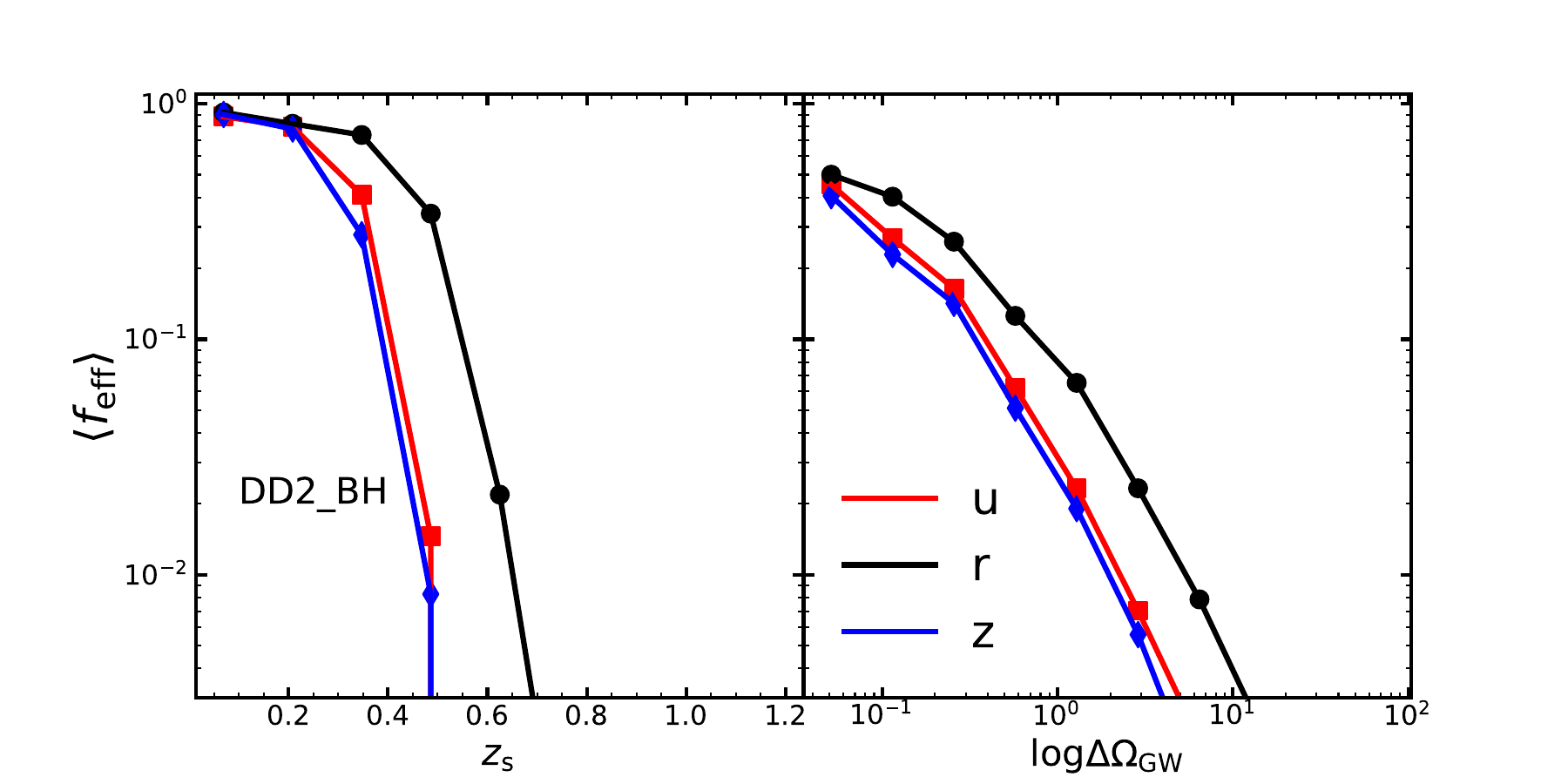}
\includegraphics[width=1.2\columnwidth]{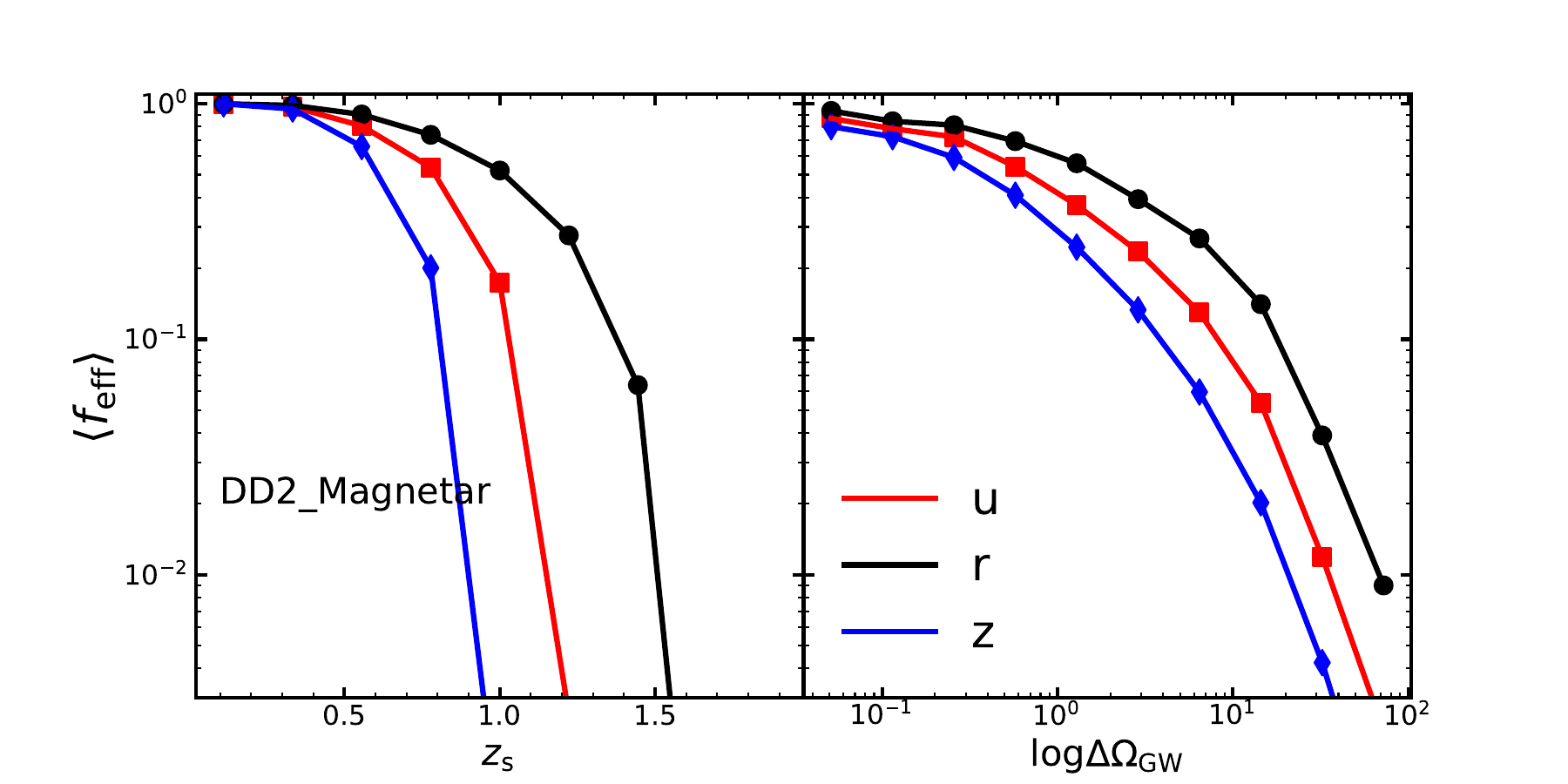}
\caption{
The distribution of the average ToO detection efficiency, $\langle f_{\rm eff} \rangle$, over BNS mergers associated with observable merger-nova signals for both BH (upper panels) and {magnetar} (lower panels) remnants, assuming that the EOS is DD2. The left and right columns display the dependency of $\langle f_{\rm eff} \rangle$ on source redshift $z_{\rm s}$ and GW localization precision $\Omega_{\rm GW}$ within a small redshift bin. In all panels, the blue, orange, and green colors represent the results for the $u$, $r$, and $z$ filters of CSST, respectively.
}
\label{fig:feff_dd2}
\end{figure*}

\begin{figure*}
\centering
\includegraphics[width=1.2\columnwidth]{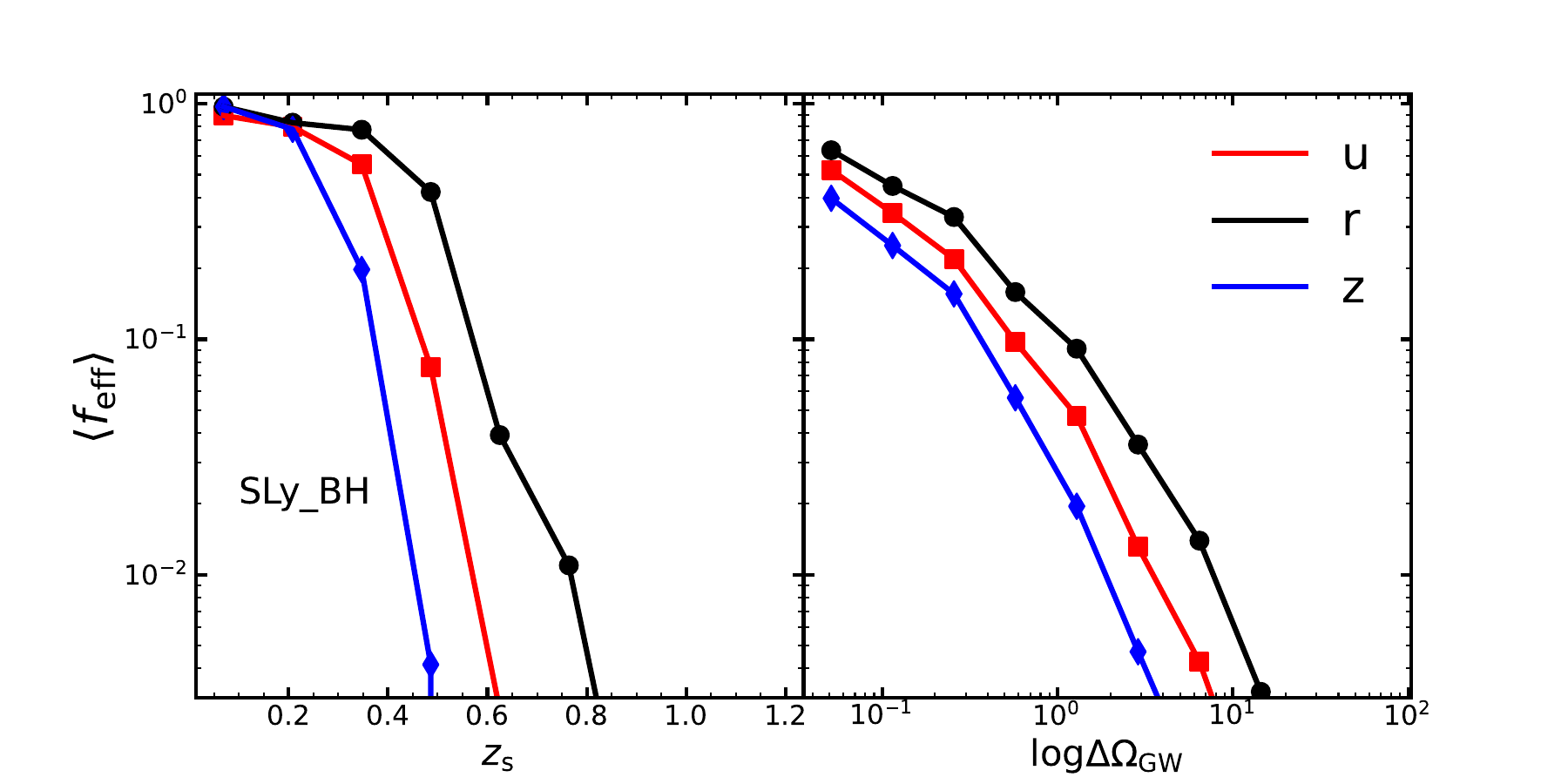}
\includegraphics[width=1.2\columnwidth]{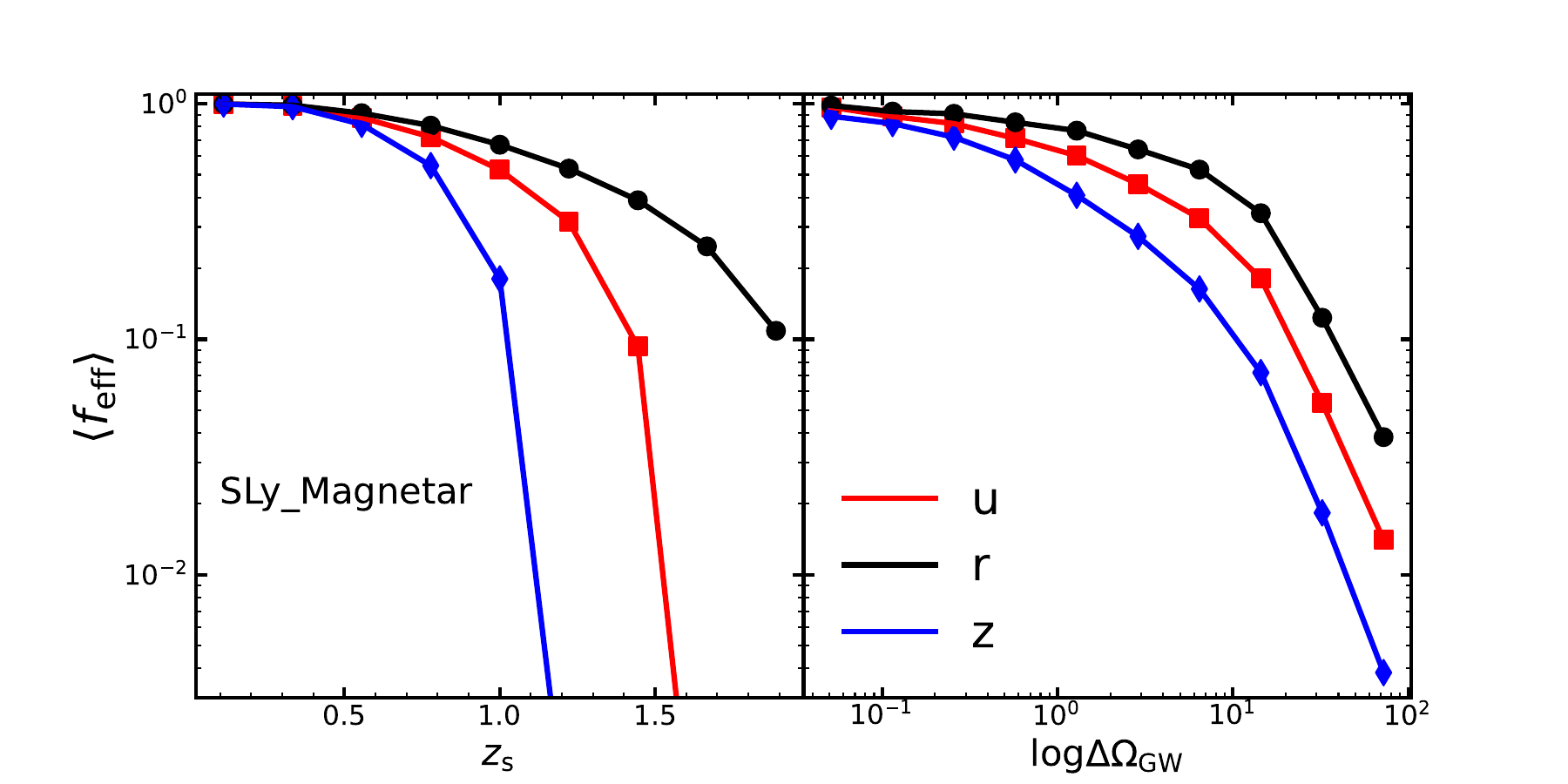}
\caption{
The legend is the same as in Figure~\ref{fig:feff_dd2}, except that the EOS is SLy.
}
\label{fig:feff_sly}
\end{figure*}

\begin{figure*}
\centering
\includegraphics[width=0.75\columnwidth]{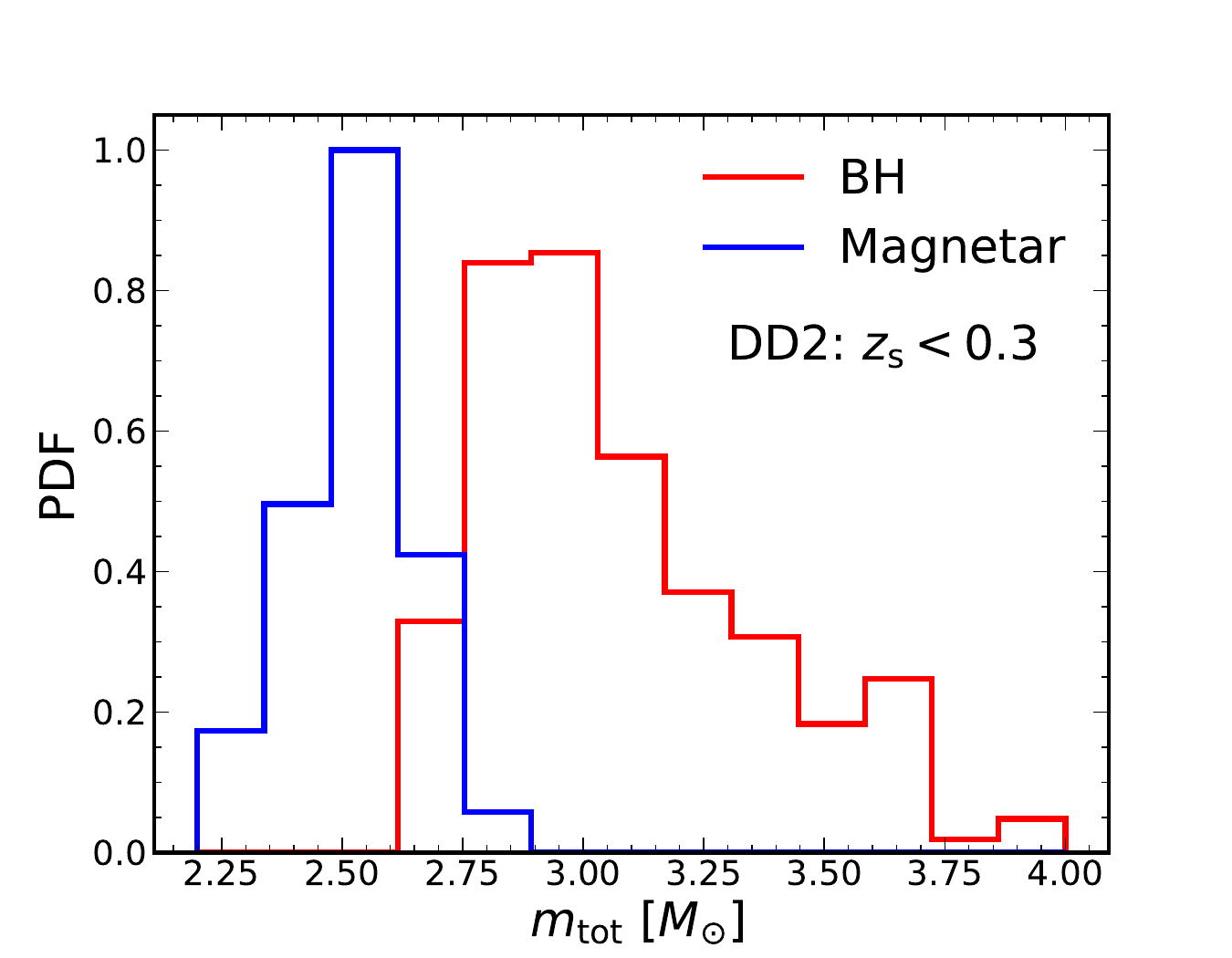}
\includegraphics[width=0.75\columnwidth]{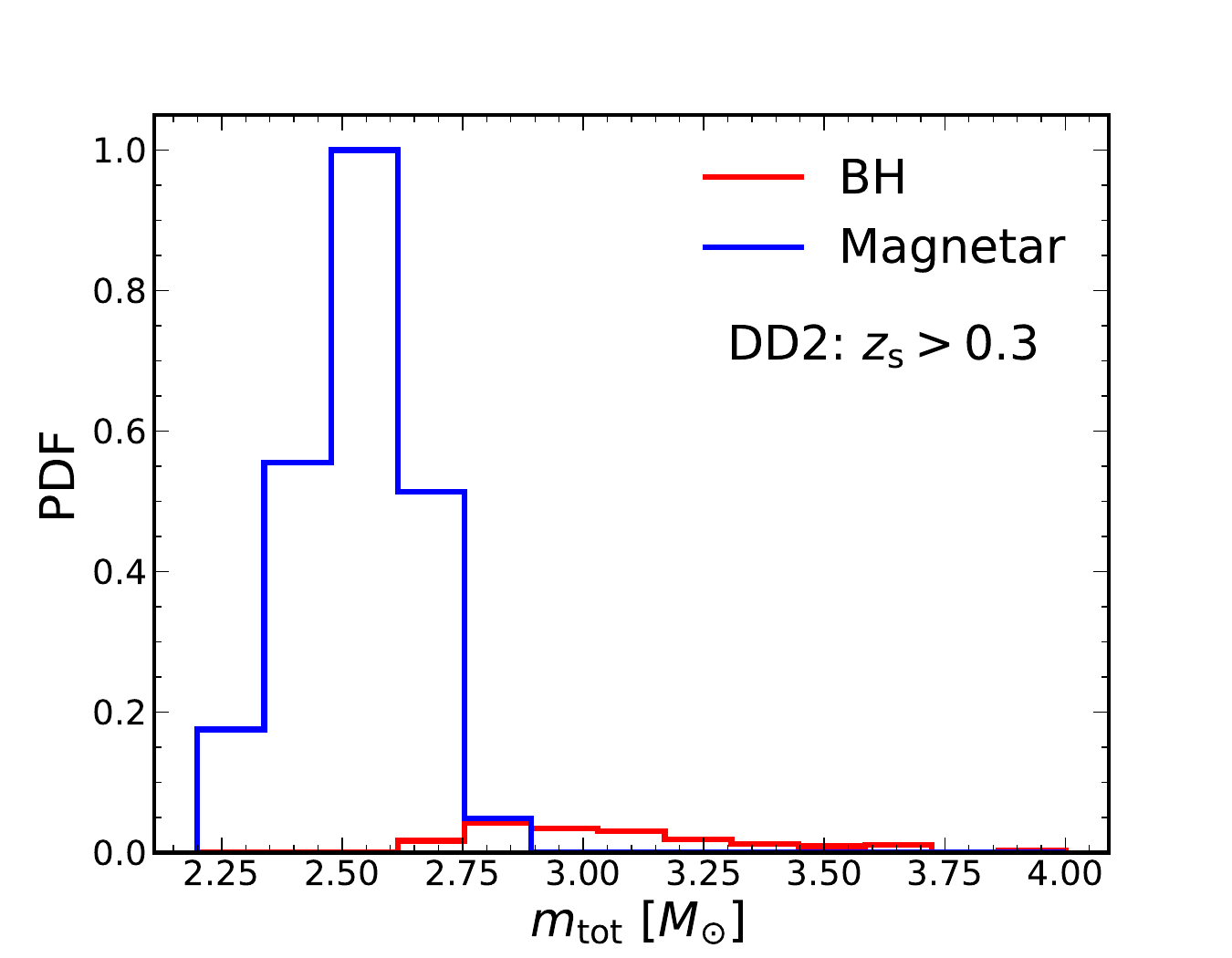}
\includegraphics[width=0.75\columnwidth]{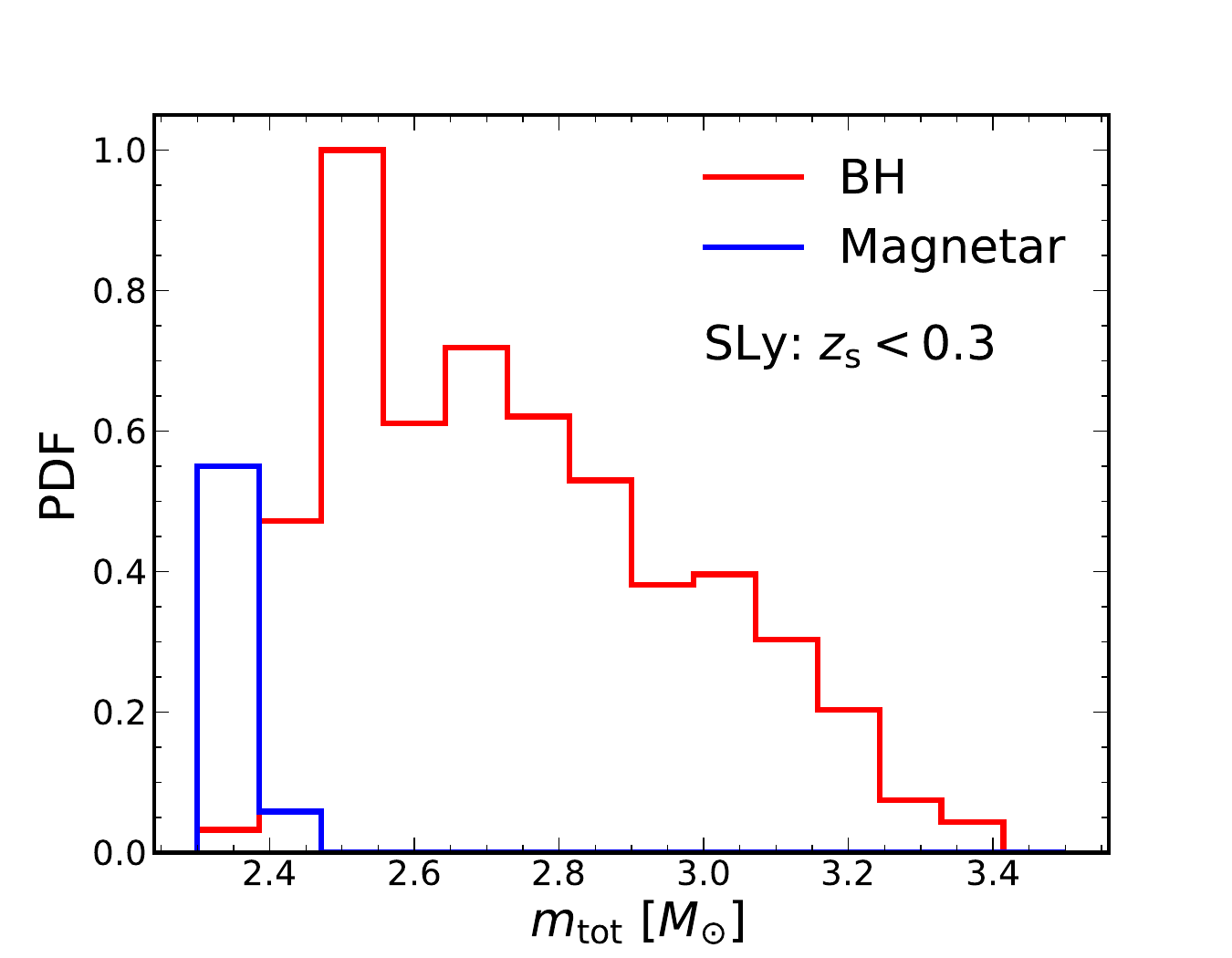}
\includegraphics[width=0.75\columnwidth]{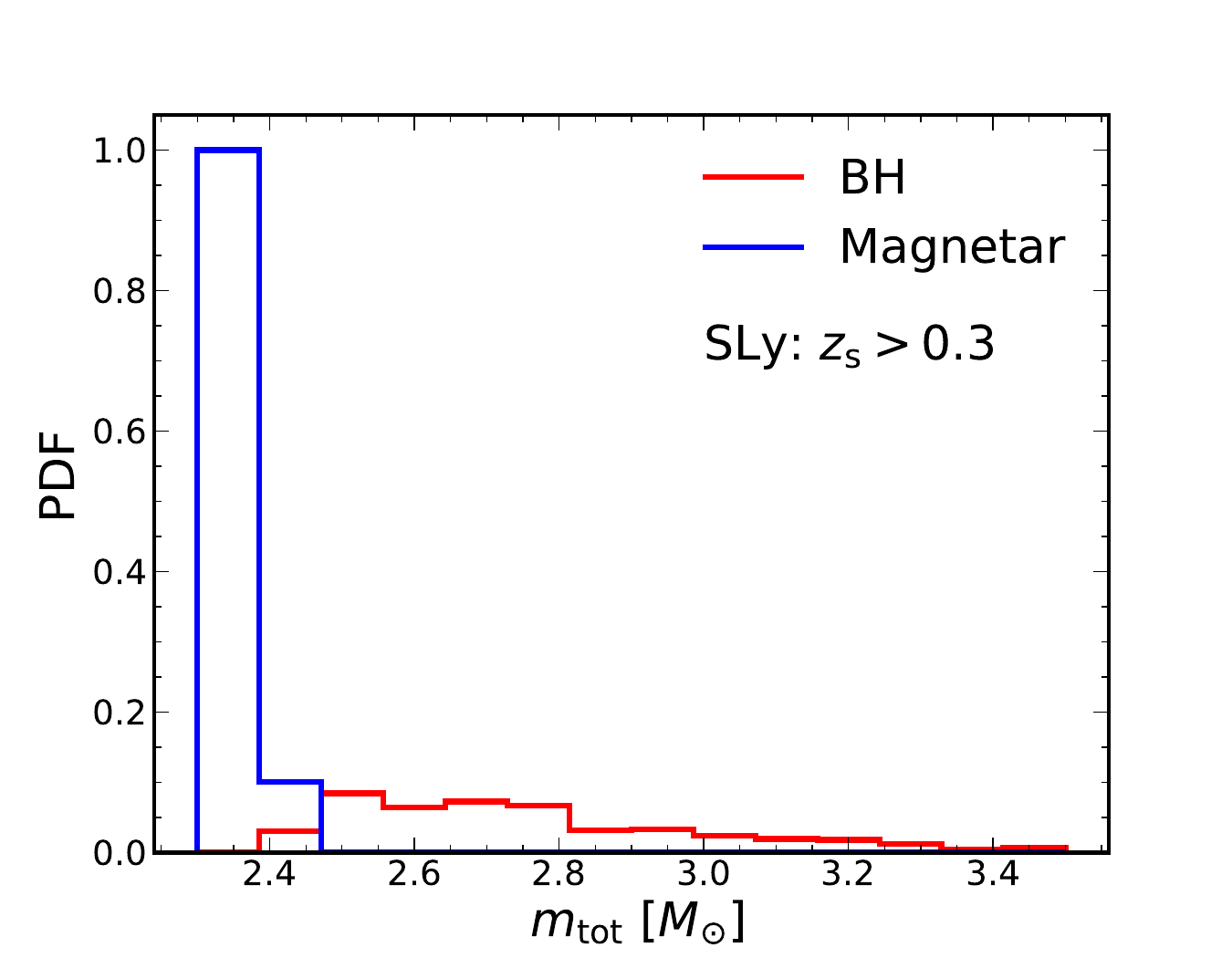}
\caption{
The probability distribution function of the total mass, $m_{\rm tot}$, for the mock BNS mergers with $z_{\rm s}<0.3$ (left panels) and $z_{\rm s}>0.3$ (right panels) , re-weighted by the ToO detection probability $P$ in the $u$ band of CSST. The top and bottom panels show the results assuming the DD2 and SLy EOS, respectively. In both panels, the red and blue lines in both panels represent the results for BH and {magnetar} remnants, respectively. 
}
\label{fig:mtot}
\end{figure*}

When BNS mergers involve {magnetar} remnants, the resulting distributions differ significantly from those without magnetar remnants due to additional energy injection from the magnetic wind. Figure~\ref{fig:dd2_smns} illustrates the normalized probability distribution of BNS mergers with {magnetar} remnant associated with merger-nova signals. In particular, the fraction of merger-nova signals possessing $\Delta T>0$ is significantly increased, i.e. about $\sim 4.61/17.4/8.87\%$, indicating that the average detection efficiency $\langle f_{\rm eff} \rangle$ will be much higher than the case of the BH remnant.  This can be easily attributed to the enhanced merger-nova signals caused by the presence of magnetic wind injection. Due to intrinsically brighter merger-nova signals, BNS mergers with {magnetar} remnants can be detected at much higher redshifts. Consequently, the distribution of their localization area peaks at larger $\Delta \Omega_{\rm GW}$ compared to those with BH remnants, as $\Delta \Omega_{\rm GW}$ strongly correlates with redshift. In the $u$, $r$ and $z$ band, the median value of $\Delta\Omega_{\rm GW}$ is about $\sim 3.70/5.91/2.40$\,deg$^2$ with standard deviation of $\sigma \sim 0.63/0.62/0.63$ dex, resulting from the selection bias of the different limiting magnitudes in various filters. In conclusion, the additional injection of energy from the magnetar remnant results in a higher fraction of merger-nova signals that possess $\Delta T>0$ and relatively larger localization areas compared to those with BH remnants.

We also show the normalized distribution of BNS mergers associated with merger-nova signals adopting a different EOS, i.e. SLy, in Figures~\ref{fig:sly_BH} and ~\ref{fig:sly_smns} for the BH and {magnetar} remnant scenarios, respectively. It can be seen that the overall results are similar with those assuming the EOS to be DD2, though the detailed numbers may vary. Again, in the remnant scenario of the {magnetar}, the fraction of merger-nova signals possessing $\Delta T>0$ is significantly larger than that of BH scenario, highlighting the importance of the magnetic wind of the magnetar.

\subsection{ToO detection efficiency 
$\langle f_{\rm eff} \rangle$ }

With the above distribution, we can estimate the average ToO detection efficiency $\langle f_{\rm eff} \rangle$ of BNS mergers associated with merger-nova signals with CSST filters. Figure~\ref{fig:feff_dd2} shows the distribution of the average $\langle f_{\rm eff} \rangle$ for both the remnants of BH (top panels) and {magnetar} (bottom panels), assuming the EOS to be DD2. The left and right columns show the dependence of $\langle f_{\rm eff} \rangle$ on source redshift $z_{\rm s}$ and GW localization precision $\Omega_{\rm GW}$ within a small redshift bin. In all panels, the blue, orange, and green colors represent the results for the $u$, $r$, and $z$ CSST filters. From this figure, we can reach the following conclusions. {First, $\langle f_{\rm eff} \rangle$ for the case of BH remnants decreases rapidly with redshift $z_{\rm s}$ and becomes less than $0.01$ at $z_{\rm s}\gtrsim 0.4-0.7$. As for the {magnetar} remnant case, this trend is much weaker and $\langle f_{\rm eff} \rangle$ can remain larger than $0.1$ until {a  high redshift $z_{\rm } \sim 1-1.5$}, due to the significantly enhanced luminosities by magnetars. In addition, $\langle f_{\rm eff} \rangle$ may be different for different bands at a given $z_{\rm s}$. For the {magnetar} case, for example, {merger-nova signals within redshift of $\sim 0.7$, $0.9$, and $1.2$ can be detected} with $\langle f_{\rm eff} \rangle$ larger than $0.3$ for the $u$, $\boldsymbol{r}$ and $z$ bands, respectively. This is mainly due to their different limiting magnitude. Second, we can see that the higher the localization precision $\Delta\Omega_{\rm GW}$, the lower the average detection efficiency $\langle f_{\rm eff} \rangle$. For the same filter at the same $\Delta\Omega_{\rm GW}$, $\langle f_{\rm eff} \rangle$ for the {magnetar} case will be substantially larger than that for a BH case. Specifically, when the localization precision is about $\Delta\Omega_{\rm GW}\sim 1 \rm deg^2$, $\langle f_{\rm eff} \rangle$ for the $u$, $r$, $z$ bands are about $0.04$, $0.09$, and $0.03$ for the BH case, {and $0.44$, $0.61$, and $0.31$ for the magnetar cases}, respectively. This deviation can be explained by the brighter light curve of merger-nova with a {magnetar} remnant, and therefore more sources can be observed with a large probability.  } 

We also show the distribution of the average $\langle f_{\rm eff} \rangle$ assuming that the EOS is SLy in Figure~\ref{fig:feff_sly}. The general trend of dependency is  similar to that with DD2 EOS, but the overall average $\langle f_{\rm eff} \rangle$ is different. This is partly due to their different intrinsic physical parameters of the merger-nova models needed to fit the GW170817 merger-nova signals and therefore lead to different intrinsic luminosity functions, especially the location of the second peak. Here we conclude that for both EOSs, we can only observe merger-nova signals with BH remnant at redshift $z_{\rm s}\sim 0.5$ with $\langle f_{\rm eff} \rangle$ larger than $0.1$, while this redshift will be substantially higher for the {magnetar} remnant case.  

Using mock BNS mergers associated with merger-nova signals, we can further estimate the distribution of the total mass $m_{\rm tot}$ of BNS mergers detected by the ToO strategy by reweighting the mass spectrum with $P$ at a given redshift range. {For example, Figure~\ref{fig:mtot} shows the probability distribution of the reweighted mass $m_{\rm tot}$ in the $u$ band with redshift $z_{\rm s}<0.3$ or  $z_{\rm s}>0.3$ assuming that the EOS is DD2 (upper panels) or SLy (lower panels). The results in other bands are similar. From this figure, we can reach the following results.  Firstly, when assuming the DD2 EOS, the detection ratio of merger-nova with magnetar and BH remnants is about $\sim 0.57$ with $z_{\rm s}<0.3$. Conversely, under the SLy EOS, this ratio decreases to $\sim 0.11$, due to their different intrinsic relative ratio between magnetar and BH remnants at $t_{\rm d}=1$ day among all the BNS mergers, i.e., $f_{\rm magnetar}/f_{\rm BH}\sim 0.45$ and $0.07$ for DD2 and SLy, respectively. Secondly, the distribution of the total mass of BNS mergers with magnetar remnants assuming DD2 is more extended than that assuming SLy, due to their different $m_{\rm TOV}$.  Thirdly, as for higher redshifts $z_{\rm s}>0.3$, merger-nova signals with BH remnants are much fainter than those with magnetar remnants, due to the absence of the efficient magnetic wind energy injection, and thus their detection efficiency $f_{\rm eff}$ is relatively much smaller than that with magnetar remnants. Therefore, the observed total mass distribution is mainly dominated by magnetar-powered cases for both EOS. }

At the end of the results section, it is worth mentioning that it is quite unlikely to monitor all the BNS mergers with telescope like CSST in the third generation GW detection era. Therefore, the detection of the merger-nova signals is very limited by the total observation time.

\section{Caveats and Limitations}

{For demonstration purpose, we made several assumptions and approximations in this paper, which may affect our estimation.} In reality, there are many complexities that one may need to take into account to make a more robust investigation. 
{In this section, we discuss the possible effects of some of these assumptions and complexities in our calculations. }

In our compact binary population synthesis simulation, we assume a linear relationship between the supernova remnant mass and their progenitor mass. However, this relationship may be over-simplified. In a recent work by \citet{chu2024}, they considered different supernova explosion mechanisms, including the core-collapse supernova, electron-capture supernova, and ultra-stripped supernova. Taking into account such more detailed mechanisms, the mass spectrum of the BNS mergers will be different, especially at the low mass end. In such a case, we find that most of the BNS mergers will be concentrated within the low-mass regime and therefore change the fraction between the BH, SMNS and stable-NS remnant significantly. Especially if the EOS is assumed to be SLy, there will be a small amount of stable-NS remnant produced because the total mass is comparably smaller. 

{
Note that we adopt the $\boldsymbol{\alpha10.\rm kb\beta0.9}$ model to generate mock BNS mergers for estimating the luminosity functions of merger-nova signals with various remnants. In the model, the common envelope ejection parameter is fixed at $\alpha=10$, the natal kick velocity is assumed to be bimodal distributed, and the ratio of the total mass before and after supernova explosion is fixed at $\beta=0.9$. Though this model was demonstrated to be most compatible with the local merger rate density of BNSs given by O1-O3 GW detections and the observed Galactic BNS systems via Bayesian factor analysis \citep{2022MNRAS.509.1557C}, one may still expect substantial uncertainties on the constraints for these parameters not only because of limited available observations but only for the extremely complex physical processes involving in the evolution of binaries. 
Such uncertainties in the constraints on the model parameters lead to uncertainties in the predicted BNS mass spectrum and thus hinder precise prediction of the merger-nova luminosity functions. For example, a larger $\alpha$ may lead to a mass spectrum more concentrated to the higher mass end \citep[e.g.,][]{2015MNRAS.448.1078Y,2018MNRAS.480.2011G}, and thus results in less BNS mergers with magnetar remnants and lower heights of the first peaks of the luminosity functions. Nevertheless, with the upgraded next-generation GW detectors, it is anticipated to detect a large number of BNS mergers which will provide better measurements on the BNS merger rate as a function of component masses and its cosmic evolution, and thus much better constraints on the model parameters. Thus one could construct a more reliable BNS mass spectrum from the population synthesis model for more robust predictions of the merger-nova luminosity function.  
} 

{As for the luminosity function (LF) estimation of the merger-nova signals, it is necessary to construct a  continuous relationship between the lifetime $\tau$ and the remnant mass $m_{\rm rem}$, which is however very difficult, since HMNS/SMNS/stable NS are supported from collapse to BHs via completely different mechanism. Currently, limited by technique issues, General relativity numerical simulations can only be conducted to investigate the BNS merger remnants with lifetime less than $\sim 100 \rm ms$ \citep[e.g.,][]{PhysRevD.95.123003,PhysRevD.98.104005,Koppel_2019}. For example, \citet{10.1093/mnras/sty2531} evolve about 35 BNS merger simulations for about $20~\rm ms$ and  study on its evolution within viscous time-scale. Besides, \citet{LUCCA202033} systematically construct a $\log\bigg(\frac{m_{1}\sqrt{q}}{M_{\rm TOV}}\bigg)-\log\tau$ relationship by extrapolating various results from numerical simulations and can nicely predict the lifetime of HMNS within several tens of milliseconds. However, this extrapolation neglects the interpretation of spin-down mechanisms and $\tau$ can not approach infinity for remnant below $M_{\rm TOV}$. Therefore, in this paper, we rather instead construct such relationship by the magentic dipole and GW spin-down mechanism, calibrated by the observation of X-ray plateaus of afterglow signals and GW170817. This may introduce uncertainties to the fraction of magnetars that can survive for at least $t_{\rm d}=1$ day due to the uncertainties of the remnant mass determination.  Notably, we restrict our discussions with $t_{\rm d}=1$ day after the BNS merger remnant formation, for the reason that the typical peak time of merger-nova signal is around 1 day. If we study on LF at  $t_{\rm d}$ smaller than 1 day, the height of the first peak will be larger, for more NS remnants can survive at that time and therefore transfer their spin-down energy into the merger-nova signals. }

{When it comes to the light curve of the merger-nova, we adopt a similar model with the anisotropic multi-components model proposed by \citet{2021MNRAS.505.1661B}, within which three different kinds of mass ejection mechanism (i.e., dynamical, viscous and neutrino wind) are taken into consideration. Notably, there are alternative ways to model the LC of merger-nova signals \citep[e.g.,][]{2017Natur.551...80K,2019NatAs...3...99B,2021MNRAS.505.1661B,2023MNRAS.522..912Z,2023MNRAS.520.2558B}. Nevertheless, with different LC models, we do not expect the predicted luminosity function at $t_{\rm d}=1$ days to change significantly, since all the models should be consistent with the observation of AT2017gfo. For example, \citet{2019MNRAS.489.5037B} developed a detailed Monte-Carlo radiation transfer code POSSIS and then fit the LC of AT2017gfo with a two-component, i.e., blue Lanthanide-free and red Lanthanide-rich mass ejecta, phenomenological model. They found that the total ejecta mass is about $\sim 0.04M_{\odot}$, which is slightly larger than the values estimated in this work. Thus, adopting a model similar to that proposed in \citet{2019MNRAS.489.5037B} may predict a luminosity function generally $\sim 0.5$ mag brighter than the one predicted here, but still remain the general triple-peak feature due to the existence of various merger remnants.  
In addition, in this work, we fix some parameters fitted by the only AT2017gfo data for all our BNS mergers, which is not accurate because several parameters, such as the opacity $\kappa$, temperature $T_{f}$ of each ejecta component, may vary among the BNS mergers in the real Universe. This may lead to a quantitative correction to our estimate.} 
{Nevertheless, these ambiguities may be resolved by future more detailed radiation models constrained by the multimessenger observations of many more such events. }

{As for the localization precision of GW sources, we adopt the widely-used Fisher information matrix method to estimate it, assuming that the non-linear terms in the signals can be neglected \citep{PhysRevD.77.042001}, i.e., equivalently the GW signals reach the limit of high S/N. As for high-S/N sources, many systems show good agreement in parameter estimation between the Fisher information matrix method and the full Bayesian analysis \citep[e.g.,][]{2013PhRvD..88h4013R}. For example, \citet{PhysRevD.89.042004} make a systematic comparison of the $50\%$ credible-interval sky areas determined by the Fisher information matrix method and coherent Bayesian analysis for GW sources with S/N $\geq10$. They found that the Fisher information matrix method generally tends to overestimate $\Delta\Omega_{\rm GW}$  by a median factor of $\sim1.6$ with standard deviation in the log of the ratio $0.6$, suggesting that our estimation may be rather conservative. Though time-consuming, a more accurate localization estimation for further multi-messenger observations shall be achieved by adopting full GW waveform simulations and Bayesian reconstruction to properly address the real observing scenarios, such as the updated framework proposed by \citet{2023ApJ...958..158K} to predict the performance of the subsequent O5 run of LVK networks and end-to-end follow-up EM surveys.} 

{Regarding the detection efficiency of the ToO strategy, we assume that each event alarmed by the next-generation GW detectors will be monitored for a uniform constant observation time $t_{\rm obs}$. However, strategies for merger-nova detection significantly depend on the scientific purposes of the sky surveys and the total time allocated for ToO observation is limited. Here, we provide brief discussion on the strategies with different scientific goals. (1) If we aim to detect more merger-novas associated with massive BNS mergers like GW190425, i.e., $m_{\rm tot}\sim 3.4 M_{\odot}$ within $T_{\rm tot}$, we should focus on systems with BH remnants. Therefore, the strategy should involve observing massive systems with low redshift and precise localization precision, such as $z_{\rm s}\lesssim 0.7$ and $\Delta\Omega_{\rm GW} \lesssim 1\rm deg^2$. In these regions, the average detection efficiency is approximately $\langle f_{\rm eff}\rangle \sim 1$. (2) Merger-novae can be detected at high redshift because of their high luminosities. Detection of high redshift merger-novae associated with GW signals, as the standard sirens, may provide a unique and independent cosmological inference on dark matter and dark energy, which may be difficult to constrain by local observations.
In this case, we should focus on low mass BNS mergers, which are likely to produce magnetar remnants, such that $\langle f_{\rm eff}\rangle$ can be higher compared with BH remnant case. In addition, it is also crucial to target mergers with small localization precision $\Delta\Omega_{\rm GW} \lesssim 1\rm deg^2$ to ensure a large $\langle f_{\rm eff}\rangle$. (3) The construction of the luminosity function of the merger-nova signals may provide information on both the EOS of BNS mergers and the merger-nova radiation model. Therefore, we may focus on the accumulation of merger-nova signals at low redshift $z_{\rm s}\lesssim 0.3$ to avoid tedious observational incompleteness correction and small number statistics, since 
merger-nova signals with both magnetar and BH remnants at such small distances will possess detection efficiency $\langle f_{\rm eff}\rangle\sim 1$. 
}

\section{Conclusions}
\label{sec:con}

In this paper, we first investigate the luminosity function of the merger-nova signals produced by BNS mergers with different EOS. By utilizing the results of binary population synthesis, GR numerical simulations and several observational constraints, we find that the luminosity function of merger-nova signals ($t_{\rm d}=1$ day) may exhibit a three-peak feature. On the one hand, the first peak is much brighter than the rest two peaks, which is due to the isotropic magnetic-wind injection of the spindown energy of the fast-rotating {magnetar} remnant. However, the second and third peaks are the natural consequence of the anisotropy induced by different angular distributions of the mass ejecta in the BH remnant scenario, including the dynamical ejecta, viscous ejecta, and neutrino wind. For different EOSs of the BNS mergers, the relative location and height of the three peaks may be different because of the different faction of the remnant type and their physical properties. 

Moreover, we estimate the average target-of-opportunity (ToO) detection efficiency $\langle f_{\rm eff}\rangle$ with CSST filters by the time span $\Delta T$ above the limiting magnitude and the localization precision of the GW detection $\Delta\Omega_{\rm GW}$. Our main conclusions are summarized as follows.
\begin{itemize}
\item The median localization precision $\Delta\Omega_{\rm GW}$ of the BNS mergers associated with observable magnetar-enhanced merger-nova signals by CSST will be worse than that of BNS mergers with the BH remnant by about $5-10$ times, due to observational bias of higher-redshift events. 
\item For both EOS, we can only observe merger-nova signals with the BH remnant at redshift $z_{\rm s}\lesssim 0.5$ with average ToO detection efficiency $\langle f_{\rm eff} \rangle$ larger than $0.1$, while the redshift will be substantially higher for the {magnetar} remnant case, i.e., $z_{\rm s}\sim 1-1.5$ depending on different filters. 
\item The probability distribution of the total mass $m_{\rm tot}$ reweighted by $P$ varies with EOS and redshift. At higher redshift, i.e., $z_{\rm s}>0.3$, the total mass distribution is dominated by the magnetar case due to strong selection effect of limiting magnitude. At lower redshift, the relative peak height between magnetar and BH case is strongly dependent on the EOS.
\end{itemize}

\section*{Acknowledgements}
We thank the anonymous referee for insightful comments. We thank Professor Luciano Rezzolla for very helpful suggestions on the calculation of the NS remnant lifetime. This work is partly supported by the Strategic Priority Program of the Chinese Academy of Sciences (grant no. XDB23040100), the National Astronomical Observatory of China (grant no. E4TG660101), the Postdoctoral Fellowship Program of CPSF under Grant Number GZB20250735 (ZC), and the National Natural Science Foundation of China under grant nos.\ 12273050.

\section*{Data Availability}

The data underlying this article will be shared on reasonable request to the corresponding author.



\bibliographystyle{mnras}
\bibliography{example} 




\appendix

\section{Merger-nova model}
\label{app:model}
The internal energy $E_{\rm int}^{\prime}$ of the mass ejecta in the comoving rest framework can be written as \citep[e.g., ][]{2010ApJ...717..245K}: 
\begin{equation}
\frac{dE_{\rm int}^{\prime}}{dt}=\xi_{\rm B}D^{-2}L_{\rm sd}^{\prime}+L_{\rm ra}^{\prime}-L_{\rm e}^{\prime}-\frac{E_{\rm int}^{\prime}}{3 V^{\prime}} \frac{dV^{\prime}}{dt^{\prime}}
\label{eq:ode3}
\end{equation}
where $\xi_{\rm B}D^{-2}L'_{\rm sd}$ is the luminosity injected from the spin-down energy ($L'_{\rm sd}$) of the magnetar, $\xi_{\rm B}$ represents the energy transformation fraction, $D$ is the Doppler factor, $L_{\rm ra}^{\prime}$ is the luminosity of the r-process, and the last term is the mechanical work done by the expansion of the radiation-dominated medium. In principle, the motion of the ejecta can be solved numerically, assuming that it is an ideal gas following the Euler equation with external energy supplies.  In this paper, we argue that the motion of the mass ejecta can be approximated as adiabatic expansion, which is reasonable for the estimation of the peak luminosity of the merger-nova signals. The typical peak time of the merger-nova is about $1$ day, assuming that the rms velocity of the ejecta is $\beta_{\rm rms}=0.1$, therefore the mechanical work term can be estimated as
\begin{equation}
\frac{E_{\rm int}^{\prime}}{3 V^{\prime}} \frac{dV^{\prime}}{dt^{\prime}} 
\sim \frac{E_{\rm int}^{\prime}}{t}\sim 10^{43} \rm erg s^{-1},
\end{equation}
which is rather small compared with the other three terms on the right-hand side of the equation, normally $\sim 10^{48}-10^{51} \rm erg^{-1}$. This validates the adiabatic expansion approximation. {If the expansion is homologous, then the velocity profile can be obtained by simple arguments on the self-similar solution of the Euler equation, assuming the polytropic index $\Gamma=4/3$ \citep[see Section 2.1.1 in][]{2018MNRAS.478.3298W}, 
\begin{equation}
\frac{dm(v)}{dv}\propto \left(1-\left(\frac{v}{v_{\rm max}}\right)^2\right)^3,
\label{eq:profile}
\end{equation}
where $v_{\rm max}$ is the maximum velocity of the ejecta and is directly related to the rms velocity of the ejecta $v_{\rm rms}$ by $v_{\rm rms}=0.273v_{\rm max}$. We also note here that the above approximated expression is also validated by Smooth Particle Hydrodynamics (SPH) simulation of BNS mergers \citep[e.g., ][]{2013MNRAS.430.2585R}.} Moreover, in such an approximation, the rms velocity of the ejecta is invariant, therefore, the Doppler factor can be absorbed in the energy-transformation factor $\xi_{\rm B}$, which leads to 
\begin{equation}
L_{\rm e} = \xi_{\rm B}L_{\rm sd}+L_{\rm ra},
\end{equation}
where we omit the prime notation for the case in the observer framework. Here we summarize that in this prescription the difference between merger-nova models of the BH scenario and the {magnetar} scenario comes from the participation of the magnetic wind, which enhances the luminosity of the merger-nova signals. 

\subsection{r-process luminosity}

As normally carried out \citep[e.g., ][]{2017ApJ...850L..37P, 2017ApJ...851L..21V, 2021MNRAS.505.1661B, 2023MNRAS.522..912Z}, we treat the luminosity of the r-process as anisotropic because of the anisotropic distribution of the material ejected in the BNS merger. In this paper, we adopt a similar model with the ANI-DVN multicomponent model proposed by \citet{2021MNRAS.505.1661B}, considering three types of ejecta, including dynamical ejecta by tidal forces or shocks in the collision of the NS cores, viscous ejecta due to the viscous torque of the accretion disk, and wind ejecta driven by the neutrino wind independently, rather than introducing different color (blue, red, purple) components. For each component, the ejected material is sketched by the angular distribution of its ejected mass [$m_{\rm dyn}(\theta)$, $m_{\rm wind}(\theta)$, or $m_{\rm vis}(\theta)$], rms velocity [$v^{\rm rms}_{\rm dyn}$, $v^{\rm rms}_{\rm wind}$, or $v^{\rm rms}_{\rm vis}$], and the opacity $\kappa_{\rm dyn}(\theta)$, $\kappa_{\rm wind}(\theta)$, or $\kappa_{\rm vis}(\theta)$, with $\theta$ representing the polar angle. The angular distribution in this paper is taken the same as that described in \citet{2021MNRAS.505.1661B}, but replacing the fixed step-break angle of the dynamical ejecta $\pi/4$ by the half-opening angle $\theta_{\rm dyn}$ estimated before.

We then discretize the polar direction of the mass ejecta into several bins uniformly distributed in $\cos\theta$ and for each bin we estimate the intrinsic luminosity of the r-process $L^{ij}_{\rm ra}(t)$ according to the nuclear heating rate $\epsilon_{\rm nuc}$ for each component \citep{2017ApJ...850L..37P} as
\begin{equation}
L^{ij}_{\rm ra}(t)= \epsilon_{\rm nuc}(t) m^{ij}_{\rm ej},
\end{equation}
where $i$ and $j$ denotes the $i$-th bin ($i=1, 2, ...,10$) and $j$-th ($j=1,2,3$) ejecta component, and $\epsilon_{\rm nuc}(t)$ is expressed as \citep{2012MNRAS.426.1940K}: 
\begin{equation}
\epsilon_{\rm nuc}(t)=\epsilon_{0}\epsilon_{Y_{\rm e}}(t)\left(\frac{\epsilon_{\mathrm{th}}(t)}{0.5}\right)\left[\frac{1}{2}-\frac{1}{\pi}\arctan\left(\frac{t-1.3\rm s}{0.11\rm s}\right)\right]^{1.3},
\end{equation}
where $t$ is in units of second, $\epsilon_0$ is the energy normalization and $\epsilon_{\rm th}$ is the thermalization efficiency described as \citet{2016ApJ...829..110B}
\begin{equation}
\epsilon_{\mathrm{th}}(t)=0.36\left[\exp\left(\frac{-0.56t}{1\mathrm{day}}\right)+\frac{\ln\left(1+0.34(t/1\mathrm{day})^{0.74}\right)}{0.34(t/1\mathrm{day})^{0.74}}\right], 
\end{equation}
and $\epsilon_{Y_{\rm e}}$ is a factor in denoting the significant heating difference for neutron-rich ejecta with electron fraction $Y_{\rm e}$ \citep{2015ApJ...813....2M}, i.e., if $Y_{e}<0.25$, $\epsilon_{Y_{\rm e}}=1$; if $Y_{\rm e}\gtrsim 0.25$, 
$\epsilon_{\rm Y_e}=0.5+2.5\{1+\exp[{4(t/t_{\rm n}-1)]}\}^{-1}$,
where $t_{\rm n}= 1$\,day. 

In most semi-analytical merger-nova models \citep[e.g., ][]{2017ApJ...850...55N, 2017ApJ...850L..37P, 2017ApJ...851L..21V,2021MNRAS.505.1661B, 2023MNRAS.522..912Z}, the emergent radiation is typically approximated as blackbody radiation to derive the spectrum of the merger-nova. However, photons emitted from the radioactive decay of heavy elements are reprocessed by ejecta material, which has a significant opacity. Consequently, the effective ejecta component, $m^{i,j}_{\rm ej}$, contributing to black-body radiation, is the mass enclosed between the diffusion region and the photosphere. The diffusion velocity $v_{\rm diff}$ and the photosphere velocity $v_{\rm ph}$ can be evaluated by setting the optical depth $\tau\sim \kappa \rho \Delta l$ to $ \sim c/v_{\rm diff}$ and $\sim 2/3$, respectively, following the treatment in \citet{2020EPJA...56....8B}. Then the effective mass of the ejecta can be estimated using Equation~\eqref{eq:profile} as
\begin{equation}
m^{i,j}_{\rm ej}=m^{i,j}(>v_{\rm diff})-m^{i,j}(>v_{\rm ph}).
\end{equation}
By this approximation, we may obtain the luminosity of the r-process for each angular bin $L^{ij}_{\rm ra}(t)$ explicitly under the assumption of holomogous expansion of the mass ejecta.

\subsection{Spin-down luminosity}
\label{app:sd}

In this paper, we adopt the fitted formula from numerical simulations for long-lived magnetar remnants of BNS mergers given by  \citet{10.1093/mnras/sty2531} to estimate the period of the newborn magnetar $P_{0}$ as
\begin{equation}
P_0=\bigg[a\bigg(\frac{m_{\rm rem}}{1M_{\odot}}-2.5\bigg)+b\bigg]~\rm ms,
\label{eq:bornP}
\end{equation}
where $a$ and $b$ are the fitting coefficients dependent on the EOS of BNS mergers \citep[see Table 1 in ][]{10.1093/mnras/sty2531}.  This formula is a perfect approximation for binaries with total mass  $\gtrsim2 M_{\odot}$ \citep[e.g., ][]{10.1093/mnras/sty2531,2020GReGr..52..108B}, which is always satisfied for the magnetar remnant produced in this work. Notably, the period estimated from the above formula is less than $P_0\lesssim 1 ~\rm ms $, even if we extrapolate it for a  $m_{\rm rem} \sim 1.4 M_{\odot}$ magnetar. This value is however substantially smaller than the values determined by the fitting of afterglow plateaus of sGRB samples by considering magnetic wind energy dissipation from a $1.4 M_{\odot}$ magnetar \citep[i.e., $\sim 5-10 \rm ms$, see][]{10.1093/mnras/sts683}, i.e.,
\begin{equation}
L_{\rm EM}=9.6\times 10^{46}R_{6}^{6}B_{14}^2P_{-3}^{-4} ~\rm erg s^{-1} ,
\end{equation}
where $B_{14}$ represents the strength of the dipolar magnetic field in units of $10^{14}$\,Gauss, $P_{-3}$ denotes the magnetar period in millisecond, and $R_6$ is the magnetar radius in units of $10^6$\,cm. A possible explanation to relieve this discrepancy is that the spin-down of the newborn magnetar may be first dominated by the GW radiation driven by deformations due to a strong inner magnetic toroidal field, rather than the magnetic wind \citep[e.g.,][]{PhysRevD.88.067304,PhysRevD.93.044065,2020GReGr..52..108B}. The GW radiation luminosity can be estimated by
\begin{equation}
L_{\rm GW}=1.1\times 10^{50}\bigg(\frac{\epsilon}{0.01}\bigg)^2I_{45}^2P_{-3}^{-6} ~\rm erg s^{-1},
\end{equation}
where $I_{45}$ is the moment of inertial in units of $10^{45}$\,g\,cm$^2$ and $\epsilon$ is the ellipticity of the NS, which is allowed to be very high if there exists a strong inner toroidal magnetic field, i.e., $\epsilon\sim 0.016$ \citep[e.g.,][]{10.1111/j.1365-2966.2008.14054.x,PhysRevD.88.067304}. 
For demonstration purpose of this work, we first calculate the birth period $P_0$ of the magnetar with mass $m_{\rm rem}$ from Equation~\ref{eq:bornP} and estimate the period $P_{-3}$ when magnetic wind dominates the spin-down process, i.e.,
\begin{equation}
    P_{-3}=5.7\times 10^2 \bigg(\frac{\epsilon}{0.01}\bigg) B_{14}^{-1}R_{6}^{-3}I_{45}
\end{equation}
which is therefore almost consistent with the observation. Then we use this value to estimate the energy carried by the isotropic magnetic wind \citep[e.g., ][]{2013ApJ...776L..40Y,2015ApJ...807..163G} injected to 
the merger-nova signals \citep[e.g., ][]{2006Sci...311.1127D, 2006MNRAS.372L..19F, 2013ApJ...771L..26G, 2013ApJ...776L..40Y}, i.e., 
\begin{equation}
L_{\rm sd}=L_{\rm EM}\left(1+\frac{t}{t_{\rm EM}}\right)^{-2}
\label{eq:sd}
\end{equation}
where $t_{\rm EM}=2\times 10^5 R^{-6}_{6}B_{14}^{-2}P_{-3}^{2}$\,s is the magnetic wind dominated spin-down timescale.

\begin{figure} 
\centering
\includegraphics[width=0.95\columnwidth]{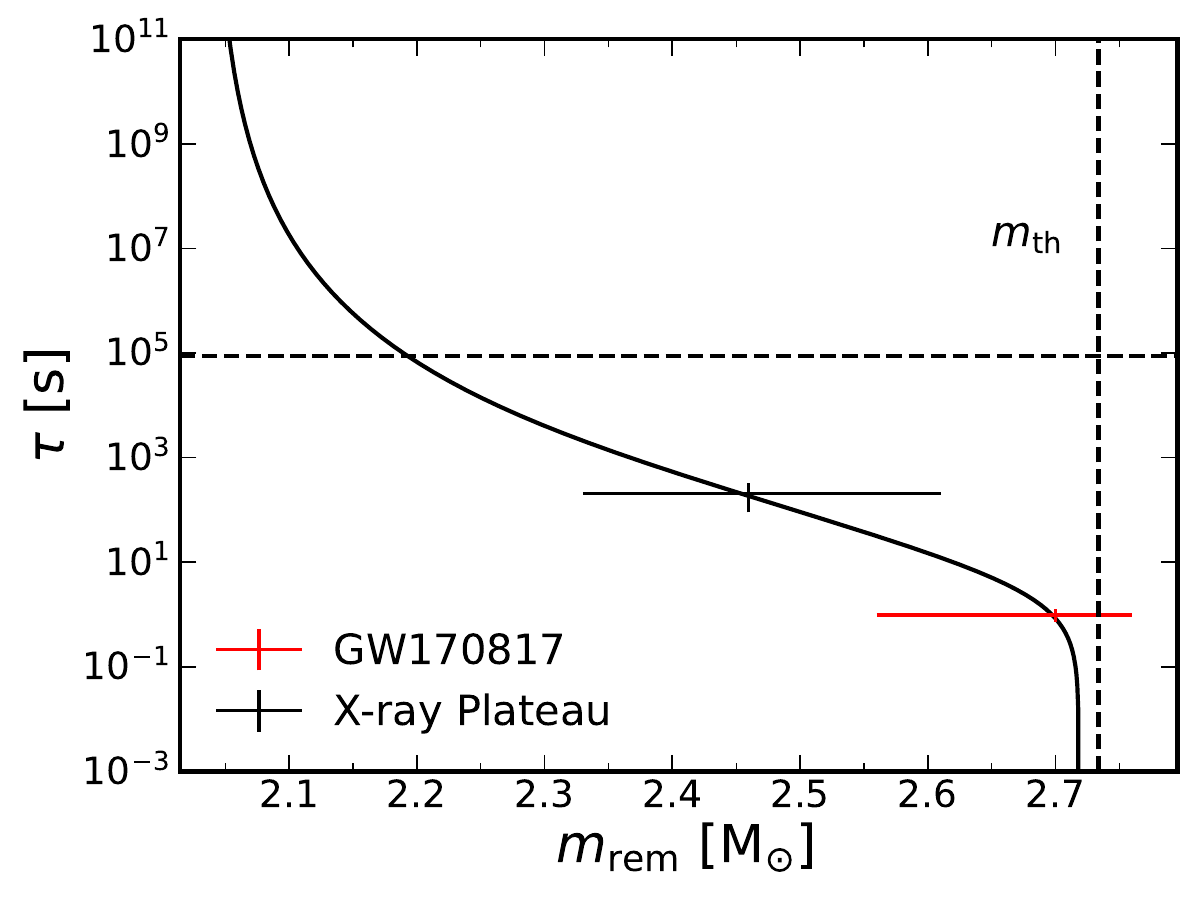}
\caption{
The lifetime $\tau$ distribution of remnant magnetar of BNS mergers with its mass $m_{\rm rem}$, assuming the EOS to be SLy. The red and black errorbars show the uncertainties for both $\tau$ and $m_{\rm rem}$ measurements on the remnant NS of GW170817 \citep[e.g.,][]{2019ApJ...876..139G} and sGRB X-ray plateaus \citep[e.g.,][]{2015ApJ...805...89L,2018ApJ...869..155S}. The dashed vertical and horizontal line shows the threshold mass for HMNS to avoid prompt collapse to BH \citep{Tootle:2021umi} and $\tau=1$ day. 
}
\label{fig:lifetime}
\end{figure}

{Once the magnetar collapses into a BH at time $\tau$ (its lifetime), the magnetic wind energy injection $L_{\rm sd}$ ceases, simultaneously terminating the enhancement of merger-nova signals. 
For a magnetar with mass smaller than $1.0 M_{\rm TOV}$, i.e., stable NS, the lifetime is infinite. Conversely, for SMNS and HMNS exceeding $1.0 M_{\rm TOV}$, the lifetime $\tau$ is dependent on the maximum spin $P_{\rm max}$ that prevent collapsing, which is directly related with $m_{\rm rem}$. Therefore, it is natural to construct a continuous distribution between $\tau$ and $m_{\rm rem}$, in which $\tau$ should approach to infinity and 0 (equivalently $P_{\rm max}$ approaches to infinity and $P_{0}$) at $m_{\rm rem }\sim M_{\rm TOV}$ and  $m_{\rm rem }\sim m_{\rm th}$ respectively, where $m_{\rm th}$ is the threshold mass for avoiding prompt collapse of HMNS remnant through differential rotation \citep[e.g.,][]{Tootle:2021umi}. In this paper, we establish such relationship between $\tau$ and $m_{\rm rem}$ using the  GW+EM spin-down mechanism discussed above \citep[e.g.,][]{2016PhRvD..93d4065G}, 
\begin{equation}
    \tau=\frac{a_{\rm GW}}{2a_{\rm EM}^2}\ln{\bigg[\bigg(\frac{a_{\rm GW}\Omega_{i}^2+a_{\rm EM}}{a_{\rm GW}\Omega_{f}^2+a_{\rm EM}}\bigg)\frac{\Omega_{f}^2}{\Omega_{i}^2}\bigg]}+\frac{\Omega_{i}^2-\Omega_{f}^2}{2a_{\rm EM}\Omega_{i}^2\Omega_{f}^2},
    \label{eq:tau}
\end{equation}
where $a_{\rm GW}=32GI\epsilon^2/5c^5$, $a_{\rm EM}=B^2R^6/6c^3I$, $\Omega_{i}=2\pi/P_0$ , $\Omega_{f}=2\pi/P_{\rm max}$ and $P_{\rm max}$ relates with $m_{\rm rem}$ through the established form \citep[e.g.,][]{2014PhRvD..89d7302L,2016PhRvD..93d4065G}: 
\begin{equation}
     \log(p_{\rm max})=k_1\log\bigg(\frac{m_{\rm rem}-M_{\rm TOV}}{M_{\rm TOV}}\bigg)+k_2
     \label{eq:pmax}
\end{equation}
where  $k_1$ and $k_2$ are fitting coefficients. In principle, $k_1$ and $k_2$  can be constrained by the observation from X-ray plateau of sGRB and GW170817. As for GW170817, the mass of the remnnant NS is about $\sim 2.7M_{\odot}$, and its lifetime $\tau$ can be constrained within $\sim 0.98~\rm s$ by its delayed jet production after $1.74~\rm s$ and kilonova signal \citep[e.g.,][]{2019ApJ...876..139G}.  As for the X-ray plateau of afterglow signals, the lifetime $\tau$ of the remnant NSs are several hundred of seconds, for example we may use a median value of $\sim 200~ \rm s$.  \citep[e.g.,][]{2015ApJ...805...89L,2018ApJ...869..155S}. Though their masses are unknown, nevertheless we can assume that $m_{\rm rem}$ is the same with the remnant mass inferred from galactic BNS systems, i.e., $m_{\rm rem}\sim 2.46 M_{\odot}$ \citep[e.g.,][]{2013arXiv1309.6635K,2025NatAs...9..552Y}.  Notably, as for the DD2 EOS, $M_{\rm TOV}$ is about $\sim 2.42 M_{\odot}$ which is very close to the remnant mass assumed for sGRBs, indicating that for stiffer EOS like DD2, only stable-NS can possess lifetime $\tau$ of several days and enhance the merger-nova signals significantly.  As for the SLy, things become different. Figure~\ref{fig:lifetime} shows the relationship between $\tau$ and $m_{\rm rem}$, where it can be seen that a small fraction of SMNS with $m_{\rm rem}\lesssim2.2 M_{\odot}$ can have  lifetime of several days, due to its small $M_{\rm TOV}$. In the context of this paper, we conclude that the magnetic wind injection for SLy and DD2 are contributed by central engine of low-mass SMNS and stable NS, respectively. }

\subsection{Flux of the merger-nova signal}

With the above two recipes, the temperature and location of the photosphere of the expanding ejecta can be directly estimated by introducing two floor temperatures, $T_{\rm f}^{\rm Ni}$ and $T_{\rm f}^{\rm LA}$ for lanthanide-free and lanthanide-rich case \citep{2021MNRAS.505.1661B} as
\begin{equation}
T^{i,j}_{\mathrm{ph}}(t)= \begin{cases}
{\left[\frac{L^{i,j}_{\mathrm{e}}(t)}{4 \pi \sigma_{\mathrm{SB}} v^{i,j 2}_{\mathrm{ph}} t^2}\right]^{\frac{1}{4}},} & \text { if } t<t^{i,j}_{\mathrm{c}}, \\ 
T^{\rm Ni/LA}_{\mathrm{f}}, & \text { if } t \geq t^{i,j}_{\mathrm{c}},
\end{cases}
\end{equation}
where $\sigma_{\rm SB}$ is the Stefan-Boltzmann constant, $v^{i,j}_{\rm ph}$ is the photosphere expansion velocity for the $j$-th ejecta in the $i$-th bin and assumed to be constant, and $t$ is the elapsed time. The radius of the photosphere is therefore 
\begin{equation}
R_{\mathrm{ph}}(t)= \begin{cases}v^{i,j}_{\mathrm{ph}} t, & \text { if } t<t^{i,j}_{\mathrm{c}}, \\ 
{\left[\frac{L^{i,j}_{\mathrm{e}}(t)}{4 \pi \sigma_{\mathrm{SB}} T_{\mathrm{f}}^4}\right]^{\frac{1}{2}},} & \text { if } t \geq t^{i,j}_{\mathrm{c}}\end{cases},
\end{equation}
where $t^{i,j}_{\rm c}=[L_{\rm e}(t)/4\pi\sigma_{\rm SB}v^{i,j 2}_{\rm ph}(T_{\rm f}^{\rm Ni/LA})^4]^{1/2}$ is the time when $T^{i,j}_{\rm ph}(t)$ reaches the flat temperature  $T_{\rm f}^{\rm Ni/LA}$ and the photosphere reaches its maximum expansion velocity.  

Thus, the spectral flux at the observer location of the approximated blackbody radiation can be eventually expressed as 
\begin{equation}
F_{\nu}(\boldsymbol{\rm n},t)=\sum_{\boldsymbol{\rm n_{ej}^{i,j}}\cdot \boldsymbol{\rm n_{ej}^{i,j}}>0}\left(\frac{R^{i,j}_{\rm ph}(t)}{D_{L}}\right)^2 B_{\nu}(T^{i,j}_{\rm ph}(t)) \boldsymbol{\rm n_{ej}^{i,j}}\cdot \boldsymbol{\rm n}\Delta \Omega^{i,j},
\label{eq:bhf}
\end{equation}
where $\boldsymbol{\rm n_{ej}^{i,j}}$ and $\boldsymbol{\rm n_{ej}^{i,j}}$ are the direction vectors of the $j$-th ejecta in the $i$-th bin and the observer, respectively, and $\Delta \Omega^{i,j}$ is the corresponding solid angle. 

Then the apparent magnitude of the merger-nova signal is easy to compute by 
\begin{equation}
m_{\rm ab}=-2.5\log F_{\nu}-48.6-5\log\left(\frac{d_{\rm L}}{10\rm Kpc}\right),
\end{equation}
where $d_{\rm L}$ is the luminosity distance of the BNS merger event.

\section{Fitting data to model}
\label{app:fit}

In this appendix, we show the details of the Bayesian fitting procedure of AT2017gfo LCs in the $u$, $g$, $r$, $i$ and $z$ bands observed by DECAM \citep[e.g.,][]{ 2017ApJ...851L..21V}. The likelihood $\mathcal{L}$ can be simply estimated by the $\chi^2$-distribution, 
\begin{equation}
\log\mathcal{L}\propto-\sum_{\alpha,\beta}\frac{1}{2}\frac{(\rm mag_{\alpha,\beta}^{\rm obs}-mag_{\alpha,\beta}^{\rm model})^{2}}{\sigma^{2}_{\alpha,\beta}+\sigma^2_{\rm sys}},
\end{equation}
where $\alpha$ marks different bands and $\beta$ marks different data points with different observation times. The systematic uncertainty $\sigma_{\rm sys}$ for the fitting is assumed to be $0.2$\,mag. The apparent magnitude of the model $\rm mag_{\alpha,\beta}^{\rm model}$ is estimated using the anisotropic 3-component merger-nova model proposed in Section~\ref{sec:remnant}. In the calculation, we discretize the polar direction into $10$ uniform bins and $100$ uniform bins in the azimuth direction, which is chosen according to the compromise between convergence and computing costs. For each EOS, i.e., SLy and DD2, we fit the LC by fixing the value of $m^{\rm dyn}_{\rm ej}$ and the $m_{\rm disk}$ inferred from equations~\ref{mass:dyn} and \ref{mass:disk} adopting the median mass of GW170817. For SLy and DD2, the inferred mass of dynamical ejecta is similar $m^{\rm dyn}_{\rm ej}=0.007/0.006M_{\odot}$ respectively. However, DD2 predicts a much more massive disk ($m_{\rm disk}=0.18M_{\odot}$) than that of SLy ($m_{\rm disk}=0.05M_{\odot}$) around the remnant of the merger. For the rest parameters, their priors are set to be uniformly distributed within the range shown in the third column of Table~\ref{table:para}, respectively.

\begin{table*}
\caption{Model parameters of the merger-nova model and their best fit values.}
\label{table:para} 
\centering
\begin{tabular}{lccccc} 
\hline

        Parameter  & Name & Prior range & SLy & DD2  \\ \hline    \hline
       $\epsilon_{0}~(\rm erg~g^{-1}~s^{-1})$ & Energy normalization &  $[0,500]\times 10^{18}$ & $183.4_{-6.0}^{+5.8}\times 10^{18}$ & $87.4_{-5.8}^{+6.4}\times 10^{18}$ \\
       $T_{f}^{\rm LA}~(\rm K)$ &  Lanthanide-rich flat temperature  & $[0,500]$  & $247.8_{-168.7}^{+172.0}$ & $252.7_{-169.5}^{+169.1}$  \\
       $T_{f}^{\rm Ni}~(\rm K)$ & Lanthanide-free flat temperature & $[3000,8000]$  & $4474.5_{-190.0}^{+205.5}$ & $4423.2_{-126.4}^{+131.7}$  \\
       $\kappa_{\rm low}~(\rm cm^2~g^{-1})$ & Low-elevation opacity & $[5,50]$ & $44.7_{-16.0}^{+5.1}$ & $39.2_{-13.1}^{+7.8}$   \\
       $\kappa_{\rm high} ~(\rm cm^2~g^{-1})$ &  High-elevation opacity & $[0,5]$ & $0.43_{-0.38}^{+1.25}$ & $1.58_{-1.07}^{+1.95}$\\
       $\kappa_{\rm wind} ~(\rm cm^2~g^{-1})$ & Wind ejecta opacity & $[0,50]$ & $33.5_{-6.7}^{+8.5}$ & $9.65_{-1.34}^{+1.71}$   \\
       $v_{\rm rms}^{\rm wind}~(\rm m~s^{-1})$ & RMS velocity of wind ejecta & $[0,0.08] c$ & $0.03_{-0.01}^{+0.01}c$ & $0.05_{-0.01}^{+0.01}c$ \\
       $\kappa_{\rm vis} ~(\rm cm^2~g^{-1})$ & Viscous ejecta opacity & $[0,50]$ &   $47.8_{-2.79}^{+1.60}$& $28.8_{-4.16}^{+4.80}$\\
       $v_{\rm rms}^{\rm vis}~(\rm m~s^{-1})$ & RMS velocity of viscous ejecta & $[0,0.08] c$ &  $0.05_{-0.00}^{+0.00}c$ & $0.07_{-0.01}^{+0.01}c$ \\
       \hline \hline
\end{tabular}
\end{table*}

\begin{figure}
\centering
\includegraphics[width=0.95\columnwidth]{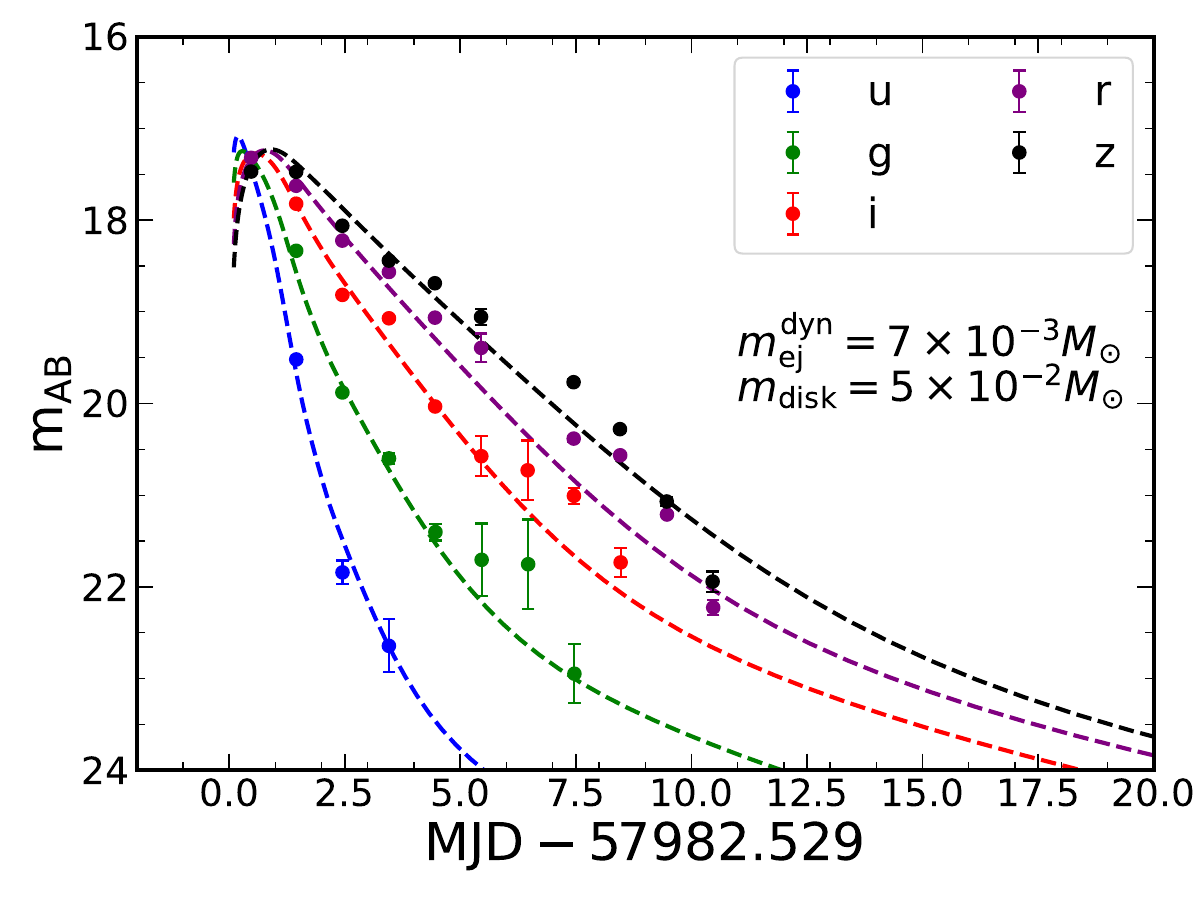}
\includegraphics[width=0.95\columnwidth]{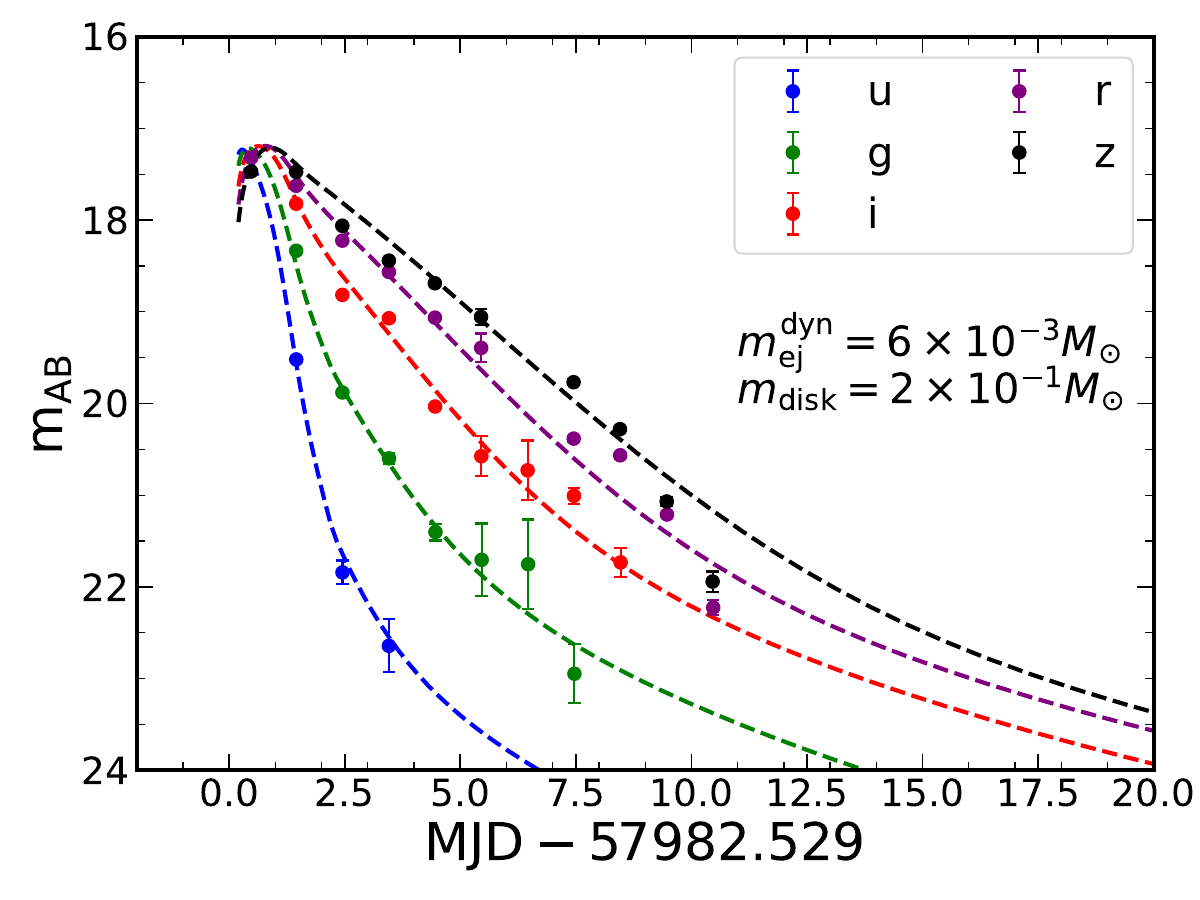}
\caption{
Multiband LCs of GW170817 kilonova AT2017gfo  and the best model fit by adopting the EOS of either SLy (top panel) or DD2 (bottom panel). Blue, green, red, purple, and black symbols represent the measurements of the $u$, $g$, $r$, $i$, and $z$ bands, respectively. The corresponding solid lines represent the best-fit results of our model.
}
\label{fig:fit_lc}
\end{figure}

Using the well-documented nested sampling \citep{2004AIPC..735..395S} Python package {Dynesty} \citep{2020MNRAS.493.3132S}, we generate the posterior distribution of the above 9 parameters utilizing more than 5000 live points with 150 threads on an AMD-EPYC-9654 processor running for about $700$ minutes. The stopping criterion is set to be when the logarithmic remaining evidence $\Delta \log\mathcal{Z}$ is less than $0.01$, which guarantees that only a very small fraction of evidence space is not explored and therefore our final results are adequate. The median values of the best fits with $16\%$ and $84\%$ errors for these parameters are listed in the fourth column of Table~\ref{table:para}. Using the median values, we present the multiband fitted LCs of AT2017gfo, adopting the EOS of both SLy and DD2, as shown in Figure~\ref{fig:fit_lc}. It is evident that in both cases the fitted LCs align with the observational data. 

\begin{figure} 
\centering
\includegraphics[width=0.95\columnwidth]{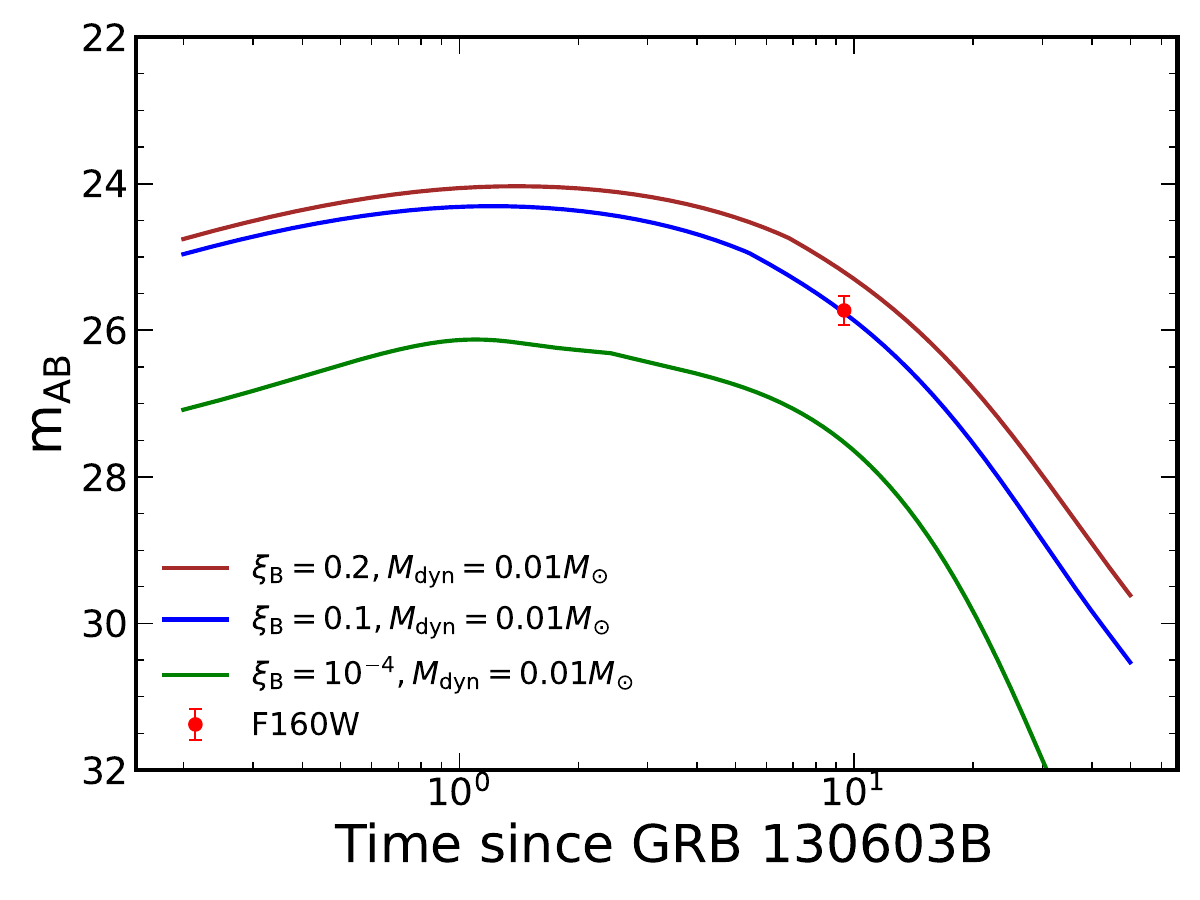}
\caption{
The predicted LC of the merger-nova associated with GRB 130603B in the F160W band, according to the {magnetar} scenario merger-nova model. {The brown and blue lines represent cases where the magnetar wind energy injection in efficient, i.e., $\xi_{\rm B}=0.2/0.1$, respectively. The green solid line represents the case where the magnetic wind is inefficient $\xi_{\rm B}=10^{-4}$, assuming $M_{\rm dyn}=0.01M_{\odot}$. The single red dot with error bars represents the observational data from HST. }
}
\label{fig:fgrb}
\end{figure}

Note that we do not constrain the value of the energy transformation factor $\xi_{\rm B}$ by the LCs of AT2017gfo, due to their extremely weak dependence on this parameter. However, we estimate $\xi_{\rm B}$ by the extreme luminous merger-nova associated with the short GRB 130603B \citep{2013Natur.500..547T}, which is often interpreted by the magnetar scenario, though with only 1 valid data point ($m_{\rm H160,\rm ab}= 25.73 \pm 0.2$). Fixing other parameters (except for $m_{\rm dyn}=0.01M_{\odot}$ to be consistent with the {magnetar} scenario) the same as in Table~\ref{table:para}, Figure~\ref{fig:fgrb} shows the resulting LCs with different $\xi_{\rm B}$ and $M_{\rm dyn}$. The red dot with error bar is the observation from the NIR F160W band of the Hubble Space Telescope (HST). We find that $\xi_{\rm B}\sim 0.2$ is adequate for explaining such a single data point by the existence of a {magnetar} remnant. 
 

\bsp	
\label{lastpage}
\end{document}